\documentclass[12pt]{article}
 \pdfoutput=1
\usepackage[table]{xcolor}
\usepackage{amsmath,amssymb,exscale}
\usepackage{graphicx}
\usepackage{epsfig}
\usepackage{multicol}
\usepackage{color}
\usepackage{mathrsfs}
\usepackage{blindtext}
 \usepackage{fancyhdr}
\usepackage{hyperref}
\usepackage{cite}
\usepackage{mathtools}
\usepackage{amsmath}
\usepackage{rotating,slashed,amsmath,charter,xcolor,catchfilebetweentags,ifluatex}

\usepackage{graphicx}
\usepackage{sidecap}

\usepackage[latin1]{inputenc} 
\usepackage{multicol}  
 
 \usepackage{titlesec}
 
 \usepackage{rotating,slashed,xcolor,amsfonts,expdlist,charter}

\numberwithin{equation}{section}

\usepackage{xcolor}
\usepackage{sectsty}

\usepackage{mdframed}
\usepackage{titletoc}

\definecolor{secnum}{RGB}{13,151,225}
\definecolor{ptcbackground}{RGB}{212,237,252}
\definecolor{ptctitle}{RGB}{0,177,235}

\titlecontents{lsection}
  [5.8em]{\sffamily}
  {\color{secnum}\contentslabel{2.3em}\normalcolor}{}
  {\titlerule*[1000pc]{.}\contentspage\\\hspace*{-5.8em}\vspace*{5pt}%
    \color{white}\rule{\dimexpr\textwidth-15.5pt\relax}{1pt}}

\usepackage{hyperref}
\hypersetup{colorlinks,bookmarksopen,bookmarksnumbered,citecolor=blus,
linkcolor=redy,pdfstartview=FitH,urlcolor=blus}
\usepackage{slashed}

\definecolor{blus}{cmyk}{1,0.9,0,0.1}
\definecolor{verdes}{cmyk}{0.99,0,0.59,0.65}
\definecolor{rossos}{cmyk}{0,1,1,0.55}
\definecolor{redy}{cmyk}{0,1,1,0.7}
\definecolor{greeny}{cmyk}{0.99,0,0.59,0.98}
\definecolor{green-go}{cmyk}{0.79,0,0.59,0.5}

\usepackage{titlesec}

\def\Lag{\mathscr{L}}

\newcommand{\beq}{\begin{equation}}
\newcommand{\eeq}{\end{equation}}

\def\hhref#1{\href{http://arxiv.org/abs/#1}{arXiv:#1}} % in bibliography

 \def\Lag{\mathscr{L}}
 
\newcommand{\tmtextbf}[1]{{\bfseries{#1}}}
\newcommand{\tmtextrm}[1]{{\rmfamily{#1}}}
\def\bp{M_{P}}

\def\be{\begin{equation}}
\def\ee{\end{equation}}
\def\ba{\begin{array} }
\newcommand{\Tr}{\,{\rm Tr}}
\def\bac{\begin{array} {c}}
\def\bacc{\begin{array} {cc}}
\def\baccc{\begin{array} {ccc}}
\def\bacccc{\begin{array} {cccc}}
\def\ea{\end{array}}
\def\bea{\begin{eqnarray}}
\def\eea{\end{eqnarray}}

\definecolor{red}{rgb}{1,0,0}

\def\psl{\hbox{\hbox{${p}$}}\kern-1.9mm{\hbox{${/}$}}}
\def\dsl{\hbox{\hbox{${\partial}$}}\kern-2.2mm{\hbox{${/}$}}}
\def\Dsl{\hbox{\hbox{${D}$}}\kern-2.6mm{\hbox{${/}$}}}

\def\Lag{\mathscr{L}}

\newcommand{\gappeq}{{\rlap{{\raise}.5ex\text{\ensuremath{>}}}{{\lower}.5ex\text{\ensuremath{\sim}}}}}
\newcommand{\lappeq}{{\rlap{{\raise}.5ex\text{\ensuremath{<}}}{{\lower}.5ex\text{\ensuremath{\sim}}}}}
\newcommand{\I}{\tmtextrm{1{\kern}-.24em l}}

\begin{document}
\topmargin -1.0cm
\oddsidemargin 0.9cm
\evensidemargin -0.5cm

{\vspace{-1cm}}
\begin{center}

\vspace{-1cm}

%TO DO
% - resubmit

 {\huge \tmtextbf{ 
\color{rossos} Natural-Scalaron Inflation}} {\vspace{.5cm}}\\
%OTHER titles

\vspace{1.9cm}

{\large  {\bf Alberto Salvio }
{\em  

\vspace{.4cm}

 Physics Department, University of Rome Tor Vergata, \\ 
via della Ricerca Scientifica, I-00133 Rome, Italy\\

\vspace{0.6cm}

I. N. F. N. -  Rome Tor Vergata,\\
via della Ricerca Scientifica, I-00133 Rome, Italy\\ 

\vspace{0.4cm}

\vspace{0.2cm}

 \vspace{0.5cm}
}

\vspace{0.2cm}

}
\vspace{0.cm}

% 
%\noindent %--------------------------------------------------------------------------------------------------------------------------------
%{\large  {\bf Alberto Salvio }
%%\vspace{.4cm}\\
%%{\large }\\
%\vspace{.3cm}
%{\em  
%
 %
%
%\vspace{0.2cm}
%
% \vspace{0.5cm}
%}

\end{center}

%
% \begin{\large abstract}
% 
\noindent --------------------------------------------------------------------------------------------------------------------------------
\begin{center}
{\bf \large Abstract}
\end{center}

\noindent A pseudo Nambu-Goldstone boson (such as an axion-like particle) is a theoretically well-motivated inflaton as it  features a naturally flat potential (natural inflation). This is because Goldstone's theorem protects its potential from sizable quantum corrections. Such corrections, however, generically generates 
an $R^2$ term in the action, which leads to another inflaton candidate because of the equivalence between the $R^2$ term and a scalar field, the scalaron, with a quasi flat potential (Starobinsky inflation). Here it is  investigated a new multifield scenario in which both the scalaron and  a pseudo Nambu-Goldstone boson are active (natural-scalaron inflation). For generality, also a non-minimal coupling is included, which is shown to emerge from microscopic theories. It is demonstrated that a robust inflationary attractor is present even when the masses of the two inflatons are comparable. Moreover, the presence of the scalaron allows to satisfy all observational bounds in a large region of the parameter space, unlike what happens in pure-natural inflation.

  \vspace{0.4cm}

\noindent --------------------------------------------------------------------------------------------------------------------------------

\vspace{1.1cm}

\vspace{2cm}

Email: alberto.salvio@roma2.infn.it

%\end{abstract}

\newpage

\tableofcontents

\vspace{0.5cm}

%\newpage

\section{Introduction}\label{Introduction}

Goldstone's theorem establishes that  a spontaneously broken continuous  symmetry corresponds to a massless scalar, known as a Nambu-Goldstone boson (NGB). The interactions of NGBs vanish at low momenta and in coordinate space are purely derivative. One can obtain small non-derivative terms in the action by adding small explicit symmetry breaking terms. In this case the NGB can acquire a potential $V$ and in particular a mass and is known as a pseudo Nambu-Goldstone boson (PNGB). 

A famous example of PNGBs is provided by the pions (or more generically the mesons), which emerge from an axial non-Abelian flavour symmetry of quantum chromodynamics (QCD); such symmetry is spontaneously broken as well as explicitly broken by the quark masses. PNGBs also appear frequently in beyond-the-Standard-Model constructions. A popular example is an axion (like) particle, namely a scalar $\phi_A$ that corresponds to a  spontaneously broken axial U(1) inexact symmetry and  can feature an interaction of the form $\sim \phi_A F_{\mu\nu}\tilde F^{\mu\nu}$ with some gauge field strength $F_{\mu\nu}$.

As pointed out in Refs.~\cite{Freese:1990rb}, a PNGB    is a theoretically well motivated inflaton as its potential  is protected  from large quantum corrections. This is because the explicit symmetry breaking terms are the only source of $V$ and can be taken arbitrarily small.  Therefore, no tuning is required to make  $V$ flat enough to be suitable for slow-roll inflation. For this reason inflation driven by a  PNGB is also known as natural inflation\footnote{For a review of axion (like) inflation see~\cite{Pajer:2013fsa}.}.

On the other hand, generically quantum corrections do generate higher curvature terms in the action $S$~\cite{Utiyama:1962sn}, the simplest\footnote{Other terms can be formed with the Ricci tensor $R_{\mu\nu}$ and the Riemann tensor $R_{\mu\nu\rho\sigma}$, for example, $R_{\mu\nu}R^{\mu\nu}$ and $R_{\mu\nu\rho\sigma}R^{\mu\nu\rho\sigma}$.} of which is $\int d^4x\sqrt{-g}\beta R^2$, where $g$ is the metric determinant and $R$ is the Ricci scalar. Indeed, neglecting the contributions of gravitons (which can be accompanied by further unknown quantum gravity effects) we find that the coefficient $\beta$ must satisfy the following renormalization group equation 
\be (4\pi)^2\frac{d\beta}{d\ln\mu}= \frac{N_V}{15}+\frac{N_F}{60}+\frac{N_S}{180} -\frac{(\delta_{ab}+6\xi_{ab})(\delta_{ab}+6\xi_{ab})}{72}, \label{running}\ee
where $\mu$ is the usual renormalization group scale and the modified minimal subtraction scheme is used. 
Here $N_V$, $N_F$, $N_S$ are the numbers of vectors, Weyl fermions and real scalars $\phi_a$ with non-minimal couplings $\xi_{ab}$ (that appear in the action as  $ S\supset \int d^4x\sqrt{-g}\xi_{ab}\phi_a\phi_b R/2$). Of course, $N_V$, $N_F$, $N_S$ cannot be zero because one must at least recover the SM  at low energies and, furthermore, the presence of a PNGB suitable for inflation also requires   additional fundamental fields. So, even neglecting the graviton  contributions, we see that setting $\beta= 0 $ is not consistent quantum mechanically (although the non-vanishing value of $\beta$ can only be inferred once the UV completion is known, see Sec.~\ref{The natural-scalaron case}). 

As originally pointed out by Starobinsky~\cite{Starobinsky:1980te} the inclusion of the $R^2$ term above also leads to a suitable inflationary scenario. This is because such term is equivalent to a scalar $z$, known as the scalaron, with a quasi-flat potential at large enough $z$. This allows to satisfy the slow-roll conditions and also leads to predictions  in agreement with the most recent cosmic microwave background (CMB) observations presented by the Planck collaboration~\cite{Ade:2015lrj}.

Motivated by this situation, in the present work we investigate a new multifield inflationary scenario, in which {\it both} a PNGB and $z$ are active during inflation. We refer to such scenario as natural-scalaron inflation. For the sake of generality,   it is also included a non-minimal coupling between the PNGB and the Ricci scalar, which can emerge, as discussed here, from quantum gravity effects. One of the main purposes of this paper is to identify the region of the parameter space and initial conditions (of the inflatons) that lead to inflationary observables in good agreement with the experimental constraints~\cite{Ade:2015lrj}. A first test of natural-scalaron inflation is whether or not this scenario features an inflationary attractor, which effectively reduces it to a quasi-single-field inflation. Indeed, this is  required by the Planck constraints on isocurvature modes. Afterwards, one should also identify the region of the parameter space and initial conditions that give us viable values of the scalar spectral index $n_s$ and the tensor-to-scalar ratio $r$. These studies are all performed in the present paper. 

The scalaron $z$ has been combined with other inflaton candidates in the literature (see e.g.~\cite{Kaneda:2015jma}). In particular, Starobinsky inflation has been combined with Higgs inflation~\cite{Bezrukov:2007ep} in~\cite{Salvio:2015kka,Salvio:2016vxi,Ema,Gundhi:2018wyz,Enckell}, while the combination with other spin-0 fields responsible for the dynamical generation of the Planck and cosmological constant scales has been explored in~\cite{Kannike:2015apa,JKubo,Salvio:2020axm}. Ref.~\cite{McDonough:2020gmn} studied the multifield inflation driven by an axion (like) particle together with the modulus of a scalar field responsible for the spontaneous breaking of the U(1) symmetry. However, the natural-scalaron inflation has never been studied before. The goal of this article  is to fill this gap and investigate where the coexistence of $z$ and an inflaton featuring a  {\it naturally} flat potential can lead us to.

This work is organised as follows. In the next section the natural-scalaron model is introduced in the Jordan frame where an $R^2$ term and a non-minimal coupling of the PNGB are present. The action is then rewritten in the Einstein frame where both scalars appear explicitly and a non-trivial field metric is present. In section~\ref{Stationary points} the stationary points (including the minima) of the Einstein-frame potential are identified and their nature is studied. In section~\ref{Slow-roll inflation and observables} the general formalism of multifield inflation is then applied to natural-scalaron inflation to investigate the presence of an inflationary attractor and to obtain its predictions for the CMB observables. Furthermore, a comparison with the constraints presented in~\cite{Ade:2015lrj} is provided. Sec.~\ref{Conclusions} contains the conclusions. 

The article also has two appendices. In appendix~\ref{microrigin} a possible microscopic origin of the Jordan frame PNGB potential and non-minimal coupling is illustrated. Appendix~\ref{Field dependent covariant masses} contains key formul\ae~that are used here to estimate the size of the isocurvature perturbations in this multifield scenario.

\section{The model}\label{model}

The part of the action responsible for inflation is (using the mostly-plus signature for the metric)
 \begin{equation} S_{\rm I} = \int d^4x\sqrt{-g}\left[\frac{F(\phi)}{2} R+\beta  R^2-\frac12 (\partial \phi)^2-V(\phi)\right]. \label{SI}\end{equation}
The parameter $\beta$  must be non-negative for stability reasons that will become clear in Sec.~\ref{Stationary points}, the  function $F$ contains the Planck mass plus a possible non-minimal coupling between $\phi$ and gravity
 %\be F(\phi)\equiv \bp^2 + \tilde F(\phi)\ee
 and must be positive in order to have a real effective Planck mass. 
 The potential
 of the PNGB $\phi$ is periodic with period $2\pi f$, where $f$ is the symmetry breaking  energy scale,~\cite{Freese:1990rb}
 \be V(\phi) = \Lambda^4 \left(1+\cos\left(\frac{\phi}{f}\right)\right) +\Lambda_{\rm cc} \label{VInf}\ee
 and $\Lambda$ is an energy scale, generically different from $f$, which corresponds to explicit symmetry breaking terms in the fundamental action. % (see Appendix~\ref{microrigin} for an explicit example).
%The potential $V$ is even  and periodic with period $2\pi f$ so we restrict ourselves to  the interval $\phi \in [0, \pi f]$. 
The constant $\Lambda_{\rm cc}$ accounts for 
%, for reasons that will become clear soon, 
the (tiny and positive) cosmological constant responsible for the observed dark energy and is completely negligible during inflation, which occurs at a much larger energy scale\footnote{A more general PNGB potential would be of the form 
 \be V(\phi) = \Lambda^4 \left(1\pm\cos\left(\frac{n\phi}{f}\right)\right)+\Lambda_{\rm cc}\ee
  with $n$ being an integer. But,  given that $V$ is even  and periodic with period $2\pi f/n$, as far as inflation is concerned we can assume~(\ref{VInf}) without lack of generality.}. 
  A possible microscopic origin of $F$ as well as $V$ is illustrated in Appendix~\ref{microrigin}. Given that $V$ is even and periodic with period $2\pi f$ we can restrict our attention to the interval 
\be\phi\in[\pi f, 2\pi f]\ee without loss of generality.

The minimum of $V$  occurs at $\phi= \pi f$, so 
\be V'(\pi f)=0. \label{min1}\ee
For simplicity, let us identify here the point of minimum $\phi= \pi f$   with today's value of $\phi$. A more general treatment will be given in Sec.~\ref{Stationary points} around Eqs.~(\ref{steq})-%,~(\ref{stz}),~(\ref{stphi}) and~
 (\ref{Ustz}).
Requiring this model to account for the observed current value of the cosmological constant, we look for a constant and homogeneous  solution of the $\phi$ field equations,
\be V'-\frac{R}{2}F' = 0. \label{varphiEOM} \ee
Today $R$ is tiny but not quite zero so we find that $V'(\pi f)=0$ implies 
\be F'(\pi f)=0. \label{min2} \ee
We can now identify
\be F(\pi f)=\bp^2 \label{Fpif}\ee
%(namely $\tilde F(\pi f)=0$)
 because  $F$ evaluated at today's value of $\phi$ is what {\it defines} the Planck mass.

 Renormalizable versions of gravity featuring in the action  four-derivative terms  of the graviton~\cite{Stelle:1976gc} (for reviews see~\cite{Salvio:2018crh,Salvio:2020axm})  favour $F\simeq \bp^2$~\cite{Salvio:2019wcp}. However, other theories of quantum gravity may lead to a different $F$, as discussed in Appendix~\ref{microrigin}. Therefore, we  keep here a generic $F$.

As well-known, the $R^2$ term  corresponds to an additional scalar. It is useful to recall here how this happens. First, one adds to the action  the  term  $$ -  \int d^4x\sqrt{-g}\,\, \beta\left( \frac{\mathcal{A}}{4\beta}-R\right)^2,$$ where $\mathcal{A}$ is an auxiliary field: indeed by using the $\mathcal{A}$ field equation one obtains immediately that this term vanishes. On the other hand, after adding that term
\begin{equation} S_{\rm I} = \int d^4x\sqrt{-g}\left[\frac{W(\phi,\mathcal{A})}{2}R -\frac{\mathcal{A}^2}{16\beta}-\frac12 (\partial \phi)^2-V(\phi)\right],  \end{equation}
where $W(\phi,\mathcal{A})\equiv F(\phi)+\mathcal{A}$.
Note that we have the non-canonical gravitational term $WR/2$. We can now go to the Einstein frame (where we have instead  the canonical Einstein term $\bp^2 R/2$)
by performing a Weyl transformation,
\be g_{\mu\nu}\rightarrow  \frac{\bp^2}{W}g_{\mu\nu},\ee
which is well-defined when $W>0$ (otherwise the transformed metric would be singular).
After performing this transformation one obtains the action in the Einstein frame\cite{Salvio:2018crh}
\begin{equation} S_{\rm I} = \int d^4x\sqrt{-g}\left[\frac{\bp^2}{2}R-\Lag_{\rm kin}-U \right], \label{Gammast}\end{equation}
where 
\be  \Lag_{\rm kin} \equiv \frac{6\bp^2}{z^2}
 \frac{(\partial \phi)^2 + (\partial z)^2}{2},\nonumber \ee
 \be U(\phi,z)\equiv  \frac{36\bp^4}{z^4}\bigg[{V(\phi)}+   \frac{1}{16\beta}\left(\frac{z^2}{6} -F(\phi)\right)^2 \bigg]\nonumber\ee
 and the new scalar $z=\sqrt{6W}>0$ has been introduced.
 
Notice that for small enough $\beta$, the Einstein-frame potential $U$ forces $z^2=6F(\phi)$ and we obtain the pure-natural inflation with a non-minimal coupling associated with $F$ described in the Einstein-frame: 
\be  \Lag_{\rm kin} = \frac12 K (\partial \phi)^2, \qquad  K = \frac{\bp^2}{F}\left(1+\frac{3F'^2}{2F}\right),\ee
\be U = \bp^4 \frac{V(\phi)}{F(\phi)^2}. \label{PureNatP}\ee 
The opposite extreme case is when $\Lambda$ is large, in which case $U$ forces $\phi$ to lie at the minimum of its potential, $\phi = \pi f$, and one recovers the pure-scalaron inflation. Recalling the way $\Lambda$ appears in the potential, Eq.~(\ref{VInf}), it is clear that an important parameter is then 
\be \rho\equiv \frac{\sqrt{\beta}\Lambda^2}{\bp^2} \label{rhodef}\ee
because $\rho\ll 1 ( \rho \gg 1)$ corresponds to pure-natural (pure-scalaron) inflation.

\section{Stationary points of the Einstein-frame potential}\label{Stationary points}
In general, the absolute minimum of  $U(\phi,z)$ is at $\phi= \pi f$ and $z=\sqrt{6F(\pi f)}=\sqrt{6}\bp$ (having neglected $\Lambda_{\rm cc}$). 

The masses associated with the fluctuations of $\{\phi,z\}$ around the minimum $\{\pi f,\sqrt{6}\bp\}$ can be obtained by diagonalizing the Hessian matrix of $V_E$ evaluated at that point in the field space. Using the expression of $V$ in~(\ref{VInf}) and the condition in~(\ref{min2}) one finds a diagonal Hessian matrix whose non-vanishing elements define the masses of the fluctuations of $\{\phi,z\}$ around the minimum
\be m_\phi = \frac{\Lambda^2}{f}, \qquad m_z =  \frac{M_P}{2\sqrt{3\beta}}. \label{mpmz} \ee
where~(\ref{Fpif}) has been used and the cosmological constant has been neglected. Here we see explicitly that the absence of tachyonic instabilities require $\beta$ to be non-negative. Given that $m_{\phi}/m_z=2\sqrt{3} \, \rho \bp/f$ the pure-natural (pure-scalaron) inflation corresponds to $m_{\phi}/m_z$ small (large).

 What is the complete set of stationary points of $U$? To answer this question in full generality we should solve the system of equations
 \be \frac{\partial U}{\partial\phi} =0, \qquad \frac{\partial U}{\partial z} =0. \label{steq}\ee
 The second equation in~(\ref{steq}) can be solved explicitly for generic $F$ and $V$ and gives
 \be z = \sqrt{6 F(\phi )+\frac{96\beta V(\phi )}{F(\phi )}}, \label{stz}\ee 
% \be z = \sqrt{\frac{6 f_0^2 F(\phi )^2+16 V(\phi )}{f_0^2 F(\phi )}},\ee 
 which, once inserted in the first equation in~(\ref{steq}), leads to
 \be V'(\phi )=\frac{2 V(\phi ) F'(\phi )}{F(\phi )}. \label{stphi}\ee
 This is an algebraic equation, which gives the value of $\phi$ at the stationary points. The corresponding values of the potential can then be computed through 
 \be U =\frac{\bp^4 V(\phi )}{F(\phi )^2+16 \beta V(\phi )}. \label{Ustz}\ee
 In the pure-natural inflation limit (small $\beta$) the first term in the denominator of the expression above dominates over the second one and one recovers the pure-natural potential in the Einstein frame, Eq.~(\ref{PureNatP}). Note that Eqs.~(\ref{steq})-%,~(\ref{stz}),~(\ref{stphi}) and~
 (\ref{Ustz}) hold in general, even if $V$ and $F$ are not periodic functions of $\phi$ and if~(\ref{min1}) and/or~(\ref{min2}) are not satisfied.
 
 Now,  using~(\ref{min1}) and~(\ref{min2}) in Eq.~(\ref{stphi}) we recover the stationary point that we have already discussed, that for a generic $\Lambda_{\rm cc}$ reads
 \be \phi=\pi f, \qquad z=\sqrt{6\bp^2+\frac{96\beta\Lambda_{\rm cc}}{\bp^2}}, \qquad U =\frac{\bp^4 \Lambda_{\rm cc}}{\bp^4+16\beta \Lambda_{\rm cc}}, \ee
 where Eqs.~(\ref{stz}) and~(\ref{Ustz}) have been used.

 Generically there could be other stationary points and even multiple minima of the potential. To illustrate this fact let us neglect the tiny $\Lambda_{\rm cc}$ and consider the PNGB potential in~(\ref{VInf}) and the simple non-minimal coupling 
\be  F(\phi) =\bp^2+\alpha \bp^2 \left(1+\cos\left(\frac{\phi}{f}\right)\right) \label{FtN} \ee
where $\alpha$ is a real parameter that must satisfy $\alpha> -1/2$ in order for the effective Planck mass to be real for all $\phi$. The $F$ in~(\ref{FtN}) is a simple choice compatible with the periodicity of $V$~\cite{Ferreira:2018nav} and the condition in~(\ref{min2}). A microscopic origin of this   non-minimal coupling is provided in Appendix~\ref{microrigin}. 
 For the choice of $F$ given in~(\ref{FtN}) Eq.~(\ref{stphi}) has the solutions 
 \be \phi_1 = \pi f, \quad  \phi_2 = 2\pi f \label{phi12} \ee	
 and, when $\alpha>1/2$,
 \be \quad \phi_3 = f \arctan\left(\frac{1-\alpha }{\alpha },\frac{\sqrt{2 \alpha -1}}{\alpha }\right),\quad \phi_4 = f \arctan\left(\frac{1-\alpha }{\alpha },-\frac{\sqrt{2 \alpha -1}}{\alpha }\right),\label{phi34}\ee 
 where $\arctan(x_1,x_2)$ gives the angle $\gamma$ (defined in the interval $[-\pi,\pi]$) whose tangent is $x_2/x_1$ taking into account where the point $\vec x \equiv(x_1 = |\vec x|\cos\gamma, x_2= |\vec x|\sin\gamma)$ is. Given the periodicity of $V$, in addition to the solutions in~(\ref{phi12}) and~(\ref{phi34}) there are  
of course all values obtained by adding integer multiples of $2\pi f$, but a part from that there are no other solutions.
The corresponding values of $z$ are dictated by Eq.~(\ref{stz}):
 \be z_1 = \sqrt{6}\bp, \quad z_2 =\sqrt{6(2 \alpha +1) \bp^2+\frac{192 \beta \Lambda ^4}{(2 \alpha +1) \bp^2}}  \ee 
 and, when $\alpha>1/2$,
 \be z_3=z_4= \sqrt{12 \bp^2+\frac{48\beta \Lambda ^4}{\alpha \bp^2}}. \ee 
 % \frac{6 f_0^2 \Lambda ^4\bp^4}{3 \bp^4 (2 \alpha  f_0+f_0)^2+16 \Lambda ^4}
 Inserting in the potential one obtains
 \be U(\phi_1,z_1) =0, \quad U(\phi_2,z_2) =\frac{2 \Lambda ^4 \bp^4}{(2 \alpha  +1)^2 \bp^4 +32\beta \Lambda ^4} \label{Ust}\ee
 and for $\alpha>1/2$
\be U(\phi_3,z_3) =U(\phi_4,z_4) =\frac{\Lambda ^4 \bp^4}{4 \alpha \bp^4+16\beta \Lambda ^4}. \ee 

The stationary point $\{\phi_1,z_1\}$ is the absolute minimum that we have already discussed. The nature of the other stationary points can be understood again by diagonalizing the Hessian matrix of $V_E$ evaluated at those stationary points. By doing so one finds that $\{\phi_3,z_3\}$ and $\{\phi_4,z_4\}$ are always saddle points, while $\{\phi_2,z_2\}$ is a saddle point for $\alpha<1/2$, but a local minimum for $\alpha>1/2$. Moreover, it is easy to see that $\phi=\phi_2$ is a point of minimum (maximum) for the pure-natural potential in the Einstein frame, Eq.~(\ref{PureNatP}), for $\alpha>1/2$ ($\alpha<1/2$).

Summarizing, for $\alpha<1/2$ the only minimum (modulo the $2\pi f$ periodicity) is the absolute minimum that we have already discussed, $\{\phi_1,z_1\}$, but for $\alpha>1/2$ there are two non-trivial minima ($\{\phi_1,z_1\}$ and $\{\phi_2,z_2\}$), one of which, $\{\phi_2,z_2\}$, has a value of $U$ (the quantity $U(\phi_2,z_2)$ given in~(\ref{Ust})) that is not negligibly small during inflation.

 \section{Multifield slow-roll inflation and observables}\label{Slow-roll inflation and observables}
  
  \subsection{General formalism}\label{General formalism}
  
 In order to derive the relevant inflationary formul\ae~it is convenient to start with a more general framework. Notice that the action in~(\ref{Gammast}) belongs to the class of multifield inflationary actions of the form
\be S_I=\int d^4x  \sqrt{|\det g|} \,\bigg[ \frac{\bp^2}{2}R  - 
\frac{K_{ij}(\Phi) }{2}\partial_\mu \phi^i\partial^\mu \phi^j-
U(\Phi)
\bigg], \label{action}
 \ee
 where $\Phi$ is an array of scalar fields with components $\phi^i$ and $K_{ij}$ is a field metric.
  For a generic function $\mathscr{F}$ of $\Phi$, we define $\mathscr{F}_{,i}\equiv \partial \mathscr{F}/\partial \phi^i$, also $\gamma^i_{jk}$ is the affine connection in the scalar field space
 \be \gamma^i_{jk}\equiv \frac{K^{il}}{2}\left(K_{lj,k}+K_{lk,j}-K_{jk,l}\right) \ee
 and $K^{ij}$ is the inverse of the field metric $K_{ij}$ (which is used to raise and lower the scalar indices $i,j,k, ...$); for example $\mathscr{F}^{,i}\equiv K^{ij}\mathscr{F}_{,j}$. The connection $\gamma^i_{jk}$ allows to define a covariant derivative in the field space: for a vector $\Phi$-dependent field  $\mathscr{V}_i$ its covariant derivative is 
 \be D_j\mathscr{V}_i\equiv \mathscr{V}_{i;j} \equiv\mathscr{V}_{i,j} - \gamma^k_{ij}\mathscr{V}_k. \ee
 
  To describe the classical part of inflation we assume the Friedmann-Robertson-Walker metric 
    \be ds^2 = a(t)^2 \left[dr^2+r^2(d\theta^2 +\sin^2\theta d\phi^2)\right]-dt^2,  \ee
    where $a$ is the cosmological scale factor and $t$ is the cosmic time. 
    In the slow-roll regime  the scalar and $a$ equations reduce to 
    \be \dot \phi^i\simeq -\frac{U^{,i}}{3H}, \qquad H^2 \simeq \frac{U}{3  \bp^2} ,   \label{slow-roll-eq}\ee
    where $H\equiv \dot a/a$ and a dot represents a derivative with respect to  $t$.
  When inflation is driven by more than one scalar field, slow-roll occurs if two conditions are satisfied (see also \cite{Chiba:2008rp} for previous studies):
\be \epsilon \equiv  \frac{  \bp^2 U_{,i}U^{,i}}{2U^2} \ll 1. \label{1st-slow-roll}\ee 
\be \left|\eta^{i}_{\,\,\, j}\right|  \ll 1, \quad \mbox{and}\quad  \left|\frac{U^{,i}}{U^{,j}}\right| = {\cal O}(1),\quad \mbox{where}\quad \eta^{i}_{\,\,\, j}\equiv \frac{\bp^2 U^{;i}_{\,\,\, ;j}}{U}. \label{2nd-slow-roll} \ee
The second condition in~(\ref{2nd-slow-roll}) is due to the fact that we have to restrict to the values of $i$ and $j$ with the smaller values of $U^{,i}$ and $U^{,j}$, because the fields with respect to which the derivatives of the potential are larger effectively do not take part in the multifield dynamics.

The equations in (\ref{slow-roll-eq}) imply the following dynamical system for $\phi^i$: 
\be \dot \phi^i=-\frac{\bp U^{,i}(\Phi)}{\sqrt{3U(\Phi)}}, \label{dynamical1}\ee
which we solve with a condition at some initial time  $t_0$: that is $\phi^i(t_0)=\phi^i_0$. Once the functions $\phi^i(t)$ are known we can obtain $H(t)$ from the second equation in (\ref{slow-roll-eq}) and introduce the number of e-folds $N$ by 
\be N(\Phi_0) \equiv \int_{t_e}^{t_0(\Phi_0)} dt' H(t'), \label{Ndef}\ee
where $t_e$ is the time when inflation ends. Dropping the label on $t_0$ and $\phi_0$ as they are generic values we have
\be N(\Phi) \equiv \int_{t_e}^{t(\Phi)} dt' H(t'). \label{Ndef2}\ee
Moreover, using 
\be \frac{dN}{dt}=H,\label{dN/dt}\ee
to express $t$ in terms of $N$ in~(\ref{dynamical1}) we obtain the slightly simpler (but equivalent) dynamical system
\be \frac{d\phi^i}{dN}=- \frac{\bp^2U^{,i}(\Phi)}{U(\Phi)}. \label{DynSys}\ee

The function of the scalar fields $N(\Phi)$ defined in~(\ref{Ndef2}) allows us to compute the curvature power spectrum $P_\mathcal{R}$, the (curvature)  scalar spectral index  $n_s$ and the tensor-to-scalar ratio $r$. Evaluating the power spectra  at horizon exit $q=a H$, the explicit formul\ae~are~\cite{Sasaki:1995aw,Chiba:2008rp}
\be P_\mathcal{R}=\left(\frac{H}{2\pi}\right)^2 N_{,i}N^{,i},\label{power-spectrum}\ee
\be  n_s =1-2\epsilon - \frac{ 2 }{ \bp^2 N_{,i}N^{,i}}+\frac{2\eta_{ij}N^{,i}N^{,j}}{N_{,k}N^{,k}},  \label{nsFormula}
\ee
\be r\equiv \frac{P_t}{P_\mathcal{R}}=\frac{8}{ \bp^2 N_{,i}N^{,i}}. \label{rW}\ee
Note that a  rescaling of the potential $U\to \lambda U$ (where $\lambda$ is a constant) rescales $P_\mathcal{R}\to \lambda P_\mathcal{R}$ but leaves $\epsilon$, $\eta^{i}_{\,\,\, j}$, $n_s$ and $r$ invariant. The invariance of $\epsilon$ and $\eta^{i}_{\,\,\, j}$ is clear from their expressions in~(\ref{1st-slow-roll}) and~(\ref{2nd-slow-roll}).  The quantities $n_s$ and $r$ are also invariant under a  rescaling of the potential  because (according to Eq.~(\ref{slow-roll-eq})) $H^2$ is proportional to $U$, thus $N\to \sqrt{\lambda} N$, and $d\phi^i/dt \to\sqrt{\lambda} d\phi^i/dt$ and so \be N_{,i} = H \frac{\partial t}{\partial \phi^i}, \qquad   N^{,i}\equiv K^{ij} N_{,j}\ee 
are invariant.

 The quantities $P_\mathcal{R}$, $n_s$ and $r$ are constrained by the results reported in~\cite{Ade:2015lrj}. The constraints on $n_s$ and $r$ are given by the solid lines in the left plot of Fig.~\ref{nsr} (or~\ref{nsr2}), where $r_{0.002}$ is the value of $r$ at the reference momentum scale $0.002~{\rm Mpc}^{-1}$, used by the Planck collaboration~\cite{Ade:2015lrj}. Regarding the curvature power spectrum,
\be P_{\cal R}(q_*) =(2.10 \pm 0.03) 10^{-9}, \label{PRobserved}\ee 
where the pivot scale  $q_* =0.05~{\rm Mpc}^{-1}$  is used as in~\cite{Ade:2015lrj}.

  \subsection{The natural-scalaron case}\label{The natural-scalaron case}
  
Let us apply now this general formalism to natural-scalaron inflation.  In that case $i=1,2$, with $\phi^1 = \phi$ and $\phi^2 =z$, and 
  \be K_{ij}(\Phi) =  \frac{6\bp^2}{z^2} \delta_{ij}. \label{KijRen}\ee
  
The properties of $P_\mathcal{R}$, $\epsilon$, $\eta^{i}_{\,\,\, j}$, $n_s$ and $r$ under rescalings of $U$ (mentioned in Sec.~\ref{General formalism}) imply that $P_\mathcal{R}$ depends linearly on $\beta$, while $\epsilon$, $\eta^{i}_{\,\,\, j}$, $n_s$ and $r$ do not depend on $\beta$ for fixed values of $\rho$ and neglecting the tiny $\Lambda_{\rm cc}$. The parameter $\beta$ can, therefore, be adjusted for each fixed $\rho$ to reproduce the observed value of $P_\mathcal{R}$ given in~(\ref{PRobserved}). 

Note that the argument around Eq.~(\ref{running}) only tells us that setting $\beta$ to zero is generically inconsistent, but does not give us a restriction on the possible non-vanishing value of this parameter. In fact, such an information can only be extracted once the UV completion is known  because it regards the high-energy value of a (running) parameter. Once the observational constraint in~(\ref{PRobserved}) is taken into account the scalaron can contribute to inflation when $\beta$ is around $10^{9}$. This high value might appear unnatural looking at Eq.~(\ref{running}), but there are examples of UV completions for which very large values of $\beta$ are theoretically favoured from the point of view of the Higgs mass naturalness in some circumstances~\cite{Kannike:2015apa}.

 In the presence of two inflatons\footnote{In the presence of $\cal N$ inflatons there are ${\cal N}-1$ independent perturbation modes orthogonal to the inflationary path. We refer to~\cite{Kaiser:2012ak} for a detail description of these modes.}  besides $P_{\cal R}$ there is another relevant scalar power spectrum corresponding to the perturbation mode orthogonal to the inflationary path, the isocurvature
 %\footnote{The correlation power spectrum $P_{\mathcal{R}I}$ is suppressed in the slow-roll approximation. }
  one $P_I(q)$.

Regarding the inflationary paths, a first thing we can note is that   Eqs.~(\ref{min1}) and~(\ref{min2}) implies that 
the line $\phi = \pi f$  in the field space is invariant under time evolution. This follows from the structure of the field equations in\footnote{The same property remains true beyond the slow-roll approximation as long as the initial conditions are assigned on the line $\phi = \pi f$ with zero ``velocity", $\dot \phi =0$. }~(\ref{DynSys}) and the fact that $\partial U/\partial \phi$ vanishes at $\phi=\pi f$. This means that whenever the initial conditions are chosen there the inflationary path occurs on that line and is identical to the one of single-field scalaron inflation. However, given the Planck data on isocurvature perturbations~\cite{Ade:2015lrj}, which constraint the ratio
\be \beta_{\rm iso} \equiv \frac{P_I}{P_{\cal R}+P_I}, \ee 
%while the correlation power spectrum $P_{{\cal R}I}$ is suppressed in the slow-roll approximation in which the equations of the curvature and isocurvature perturbations decouple~\cite{Kaiser:2012ak} (the turn rate is small during slow-roll.
it is important that the initial conditions are chosen close to an inflationary attractor. Therefore, one first has to establish the existence of such a path.

 \begin{figure}[t]
\begin{center}
 \includegraphics[scale=0.54]{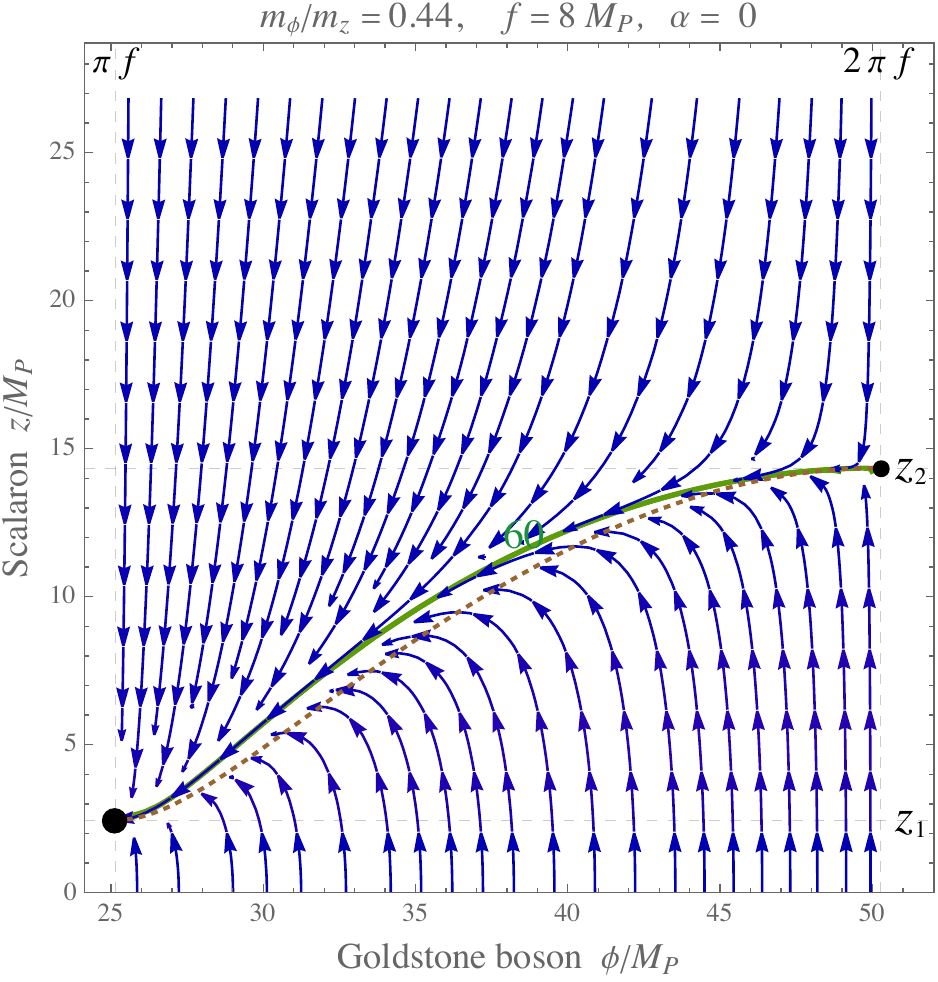}  
\hspace{0.3cm}  
\includegraphics[scale=0.54]{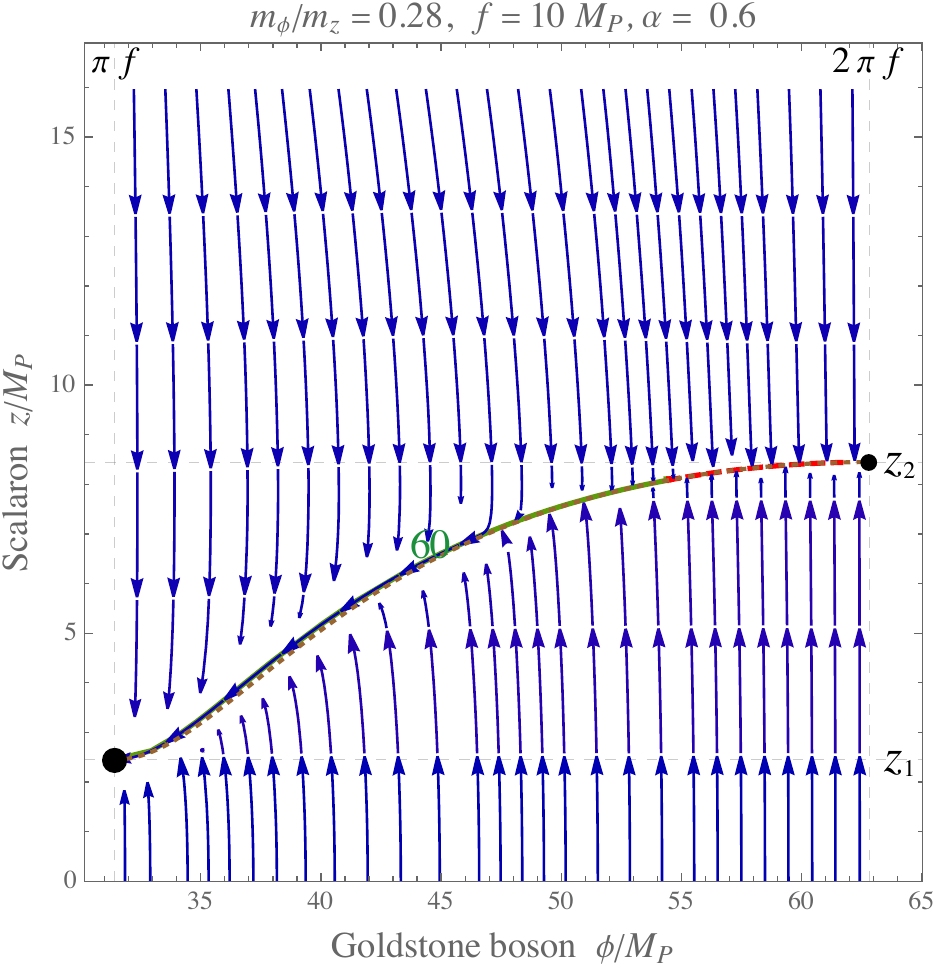} 
 \end{center}

   \caption{\em The inflationary paths in natural-scalaron inflation obtained by solving Eq.~(\ref{DynSys}) and varying the initial conditions. The green solid line highlights the inflationary attractor, while the brown dotted line represents Eq.~(\ref{stz}). The region close to the attractor corresponding to about 60 e-folds before the end of inflation is located where the corresponding number appears. The left (big) dot represents the global minimum of the potential, $\{\phi_1, z_1\}$, while the right (smaller) dot represents the stationary point $\{\phi_2, z_2\}$, which can be a saddle point (like in the left plot, without the non-minimal coupling) or a local minimum (like in the right plot). The red dashed line in the right plot allows to identify the initial conditions that are attracted to  $\{\phi_2, z_2\}$ when this point is a local minimum, all the other initial conditions form the basin of attraction of the global minimum.}
\label{attractor}
\end{figure}

 Of course, for $\rho \gg1$ or $\rho\ll 1$ we certainly have an inflationary attractor because, as already noticed, in this case one recovers pure-scalaron or pure-natural inflation, respectively. Remarkably, an inflationary attractor is also present for intermediate values of $\rho$, as shown  in Fig.~\ref{attractor} (green solid line), where $\rho\sim 1$. If the initial conditions are assigned outside this special path the scalars quickly reach it and slow-roll inflation occurs  after they have approached it. For $\rho$ small enough the attractor is well approximated by Eq.~(\ref{stz}) because in that case pure-natural inflation is a good approximation and $z$ is close to the solution of $\partial U/\partial z = 0$. However, for $\rho\gtrsim 1$ there is some sizable difference between the attractor and Eq.~(\ref{stz}), as shown in Fig.~\ref{attractor} (the left plot has $\rho\simeq 1.0$, while the right one has $\rho\simeq 0.8$). The attraction to the green solid line of Fig.~\ref{attractor} is strong enough to satisfy the Planck constraints on isocurvature perturbations. Using the formalism of Ref.~\cite{Kaiser:2012ak}, we obtain $\beta_{\rm iso} \sim 10^{-6}$  for the left plot and a much lower value for the right plot.

 In general in natural-scalaron inflation  $\beta_{\rm iso}$ is sufficiently small because in a large region of the parameter space the isocurvature perturbations have an effective mass $m^2_{ss}$ (defined in Appendix~\ref{Field dependent covariant masses}) close to the inflationary Hubble rate $H$ for a relevant number of e-folds and, therefore, the amplitude of these scalar perturbations is suppressed at superhorizon scales.

  In the left plot of Fig.~\ref{attractor} the non-minimal coupling is absent, while in the right one the non-minimal coupling is set equal to that in~(\ref{FtN}) with $\alpha=0.6>1/2$. 
 % two plots  in Fig.~\ref{attractor} have two different choices of $\alpha$ such that 
  Therefore, the stationary point $\{\phi_2,z_2\}$ discussed in Sec.~\ref{Stationary points} is a saddle point in the left plot ($\alpha=0<1/2$) and a local minimum in the right one ($\alpha>1/2$). The right plot of Fig.~\ref{attractor} also shows the basin of attraction of the global minimum $\{\phi_1,z_1\}$ when $\{\phi_2,z_2\}$ is also a minimum. This basin  is large enough to accomodate 60 e-folds  of inflation and more.

 The setups considered in Fig.~\ref{attractor} corresponds to realistic values of $n_s$ and $r$. Setting the number of e-folds to $60$, for the left plot we have 
 \be n_s(q_*) \simeq 0.967,\qquad r_{0.002} \simeq 0.0041 \ee
 while for the right plot
  \be n_s(q_*) \simeq 0.964, \qquad r_{0.002} \simeq 0.0059.\ee
   The curvature power spectrum in~(\ref{PRobserved}) can be reproduced by choosing $\beta$ appropriately, as explained above.

 \begin{figure}[t]
\begin{center}
 \includegraphics[scale=0.54]{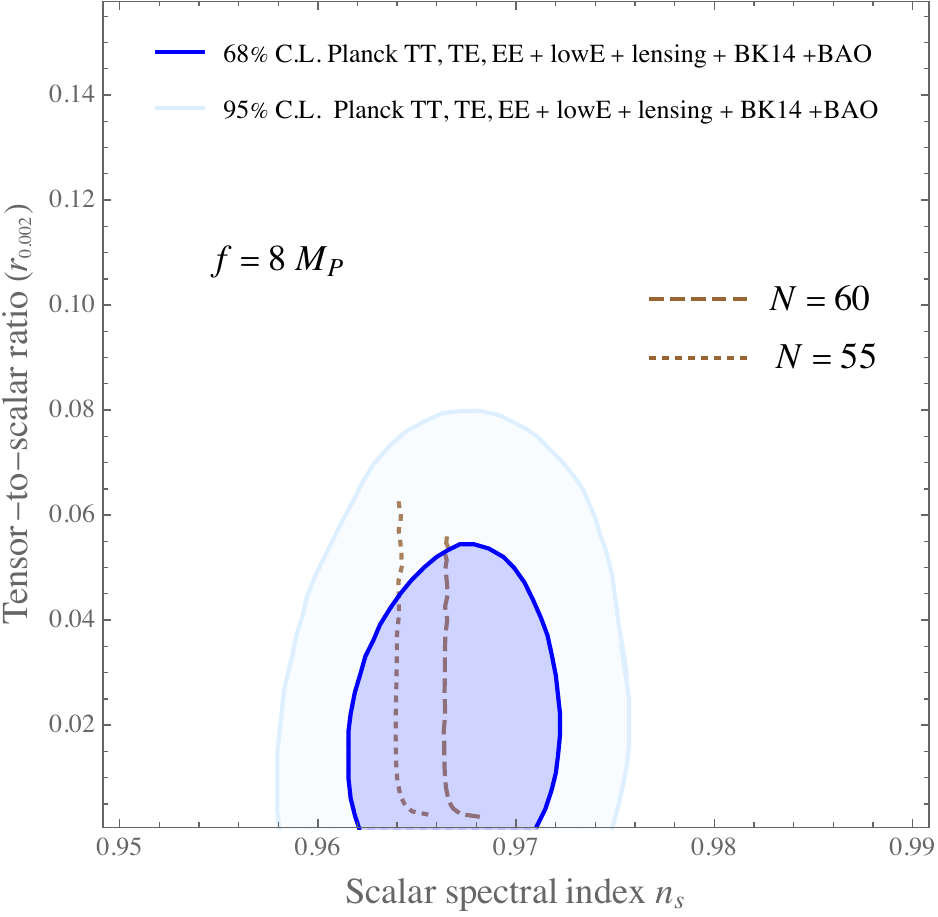}  
\hspace{0.3cm}   
\includegraphics[scale=0.54]{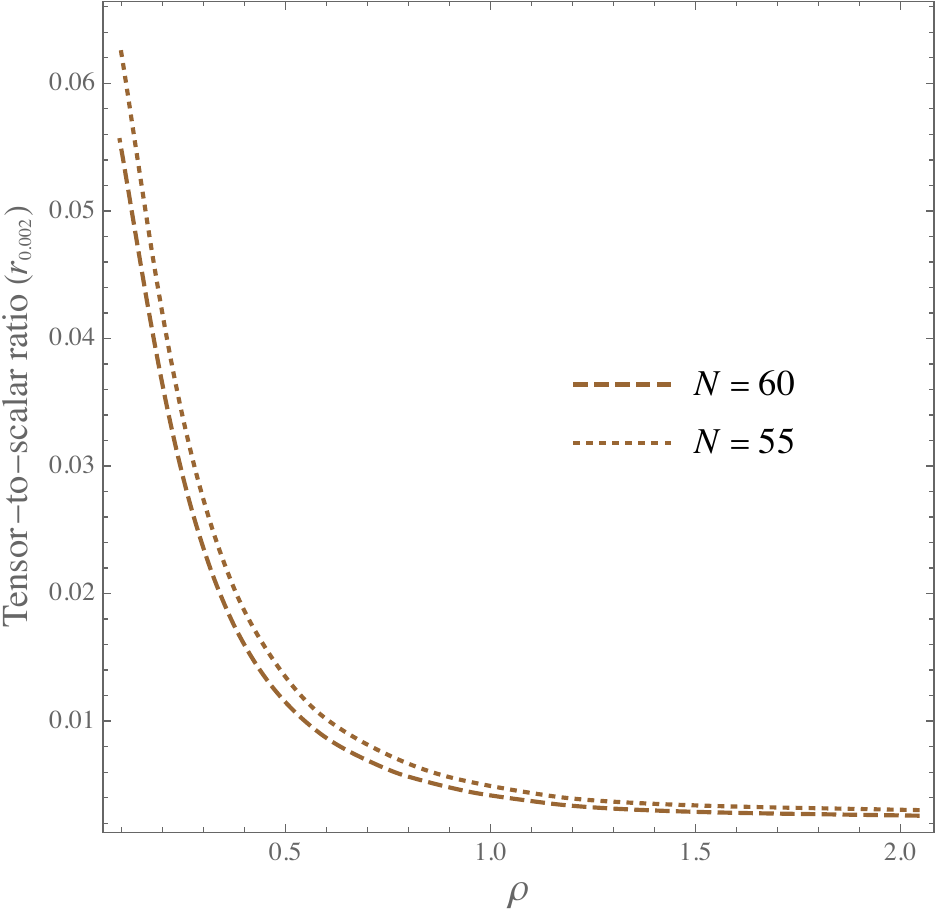} 
 \end{center}

   \caption{\em {\bf Left plot:} comparison between the theoretical predictions for $n_s$ and $r_{0.002}$ and the observational bounds in~\cite{Ade:2015lrj}.  {\bf Right plot:} corresponding variation of $r_{0.002}$ as a function of the parameter $\rho$ defined in Eq.~(\ref{rhodef}). 
}
   \vspace{0.1cm}
   {\em 

   In this figure a vanishing non-minimal coupling has been chosen, $F=\bp^2$.
   }
\label{nsr}
\end{figure}

In the left plot of Fig.~\ref{nsr} the theoretical predictions for $n_s$ and $r_{0.002}$ (varying $\rho$) from natural-scalaron inflation are compared with the observational bounds obtained combing the Planck data with the BICEP2/Keck Array (BK14) and baryon acoustic oscillation (BAO) data~\cite{Ade:2015lrj}. The right plot shows the corresponding variation of $r_{0.002}$ as a function of $\rho$. We numerically checked that the model interpolates between the pure-natural inflation predictions (small $\rho$) and the pure-scalaron inflation predictions (large $\rho$). As clear from   Fig.~\ref{nsr} there is a quite large range of values of $\rho$ such that the natural-scalaron model is in perfect agreement with the observational data at 1$\sigma$ level unlike pure-natural inflation~\cite{Stein:2021uge}. This is the case even in the absence of non-minimal couplings, which is the case in Fig.~\ref{nsr}. For all values of $\rho$ considered in Fig.~\ref{nsr} we find very small values of $\beta_{\rm iso}$ and we checked that the observational bounds on isocurvature perturbations are satisfied. The largest value of $\beta_{\rm iso}$ found is of order $10^{-4}$ and is obtained for the largest value of $\rho$ considered in Fig.~\ref{nsr}.

\begin{figure}[t]
\begin{center}
 \includegraphics[scale=0.54]{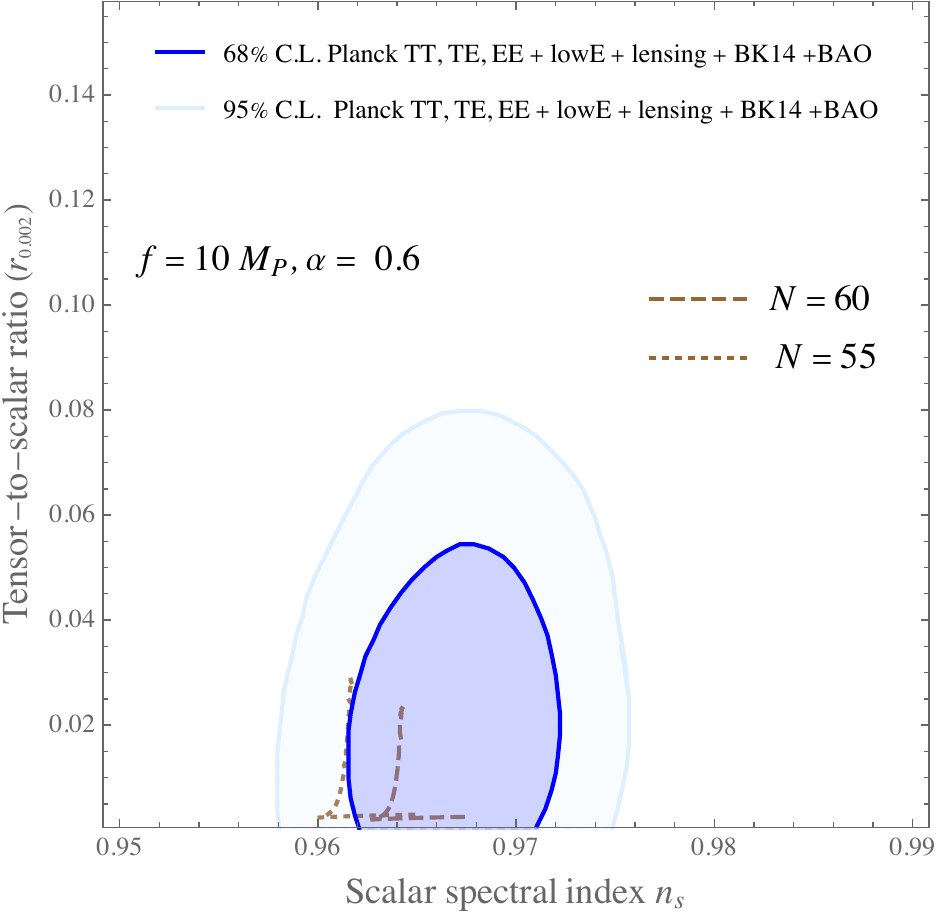}  
\hspace{0.3cm}  
\includegraphics[scale=0.54]{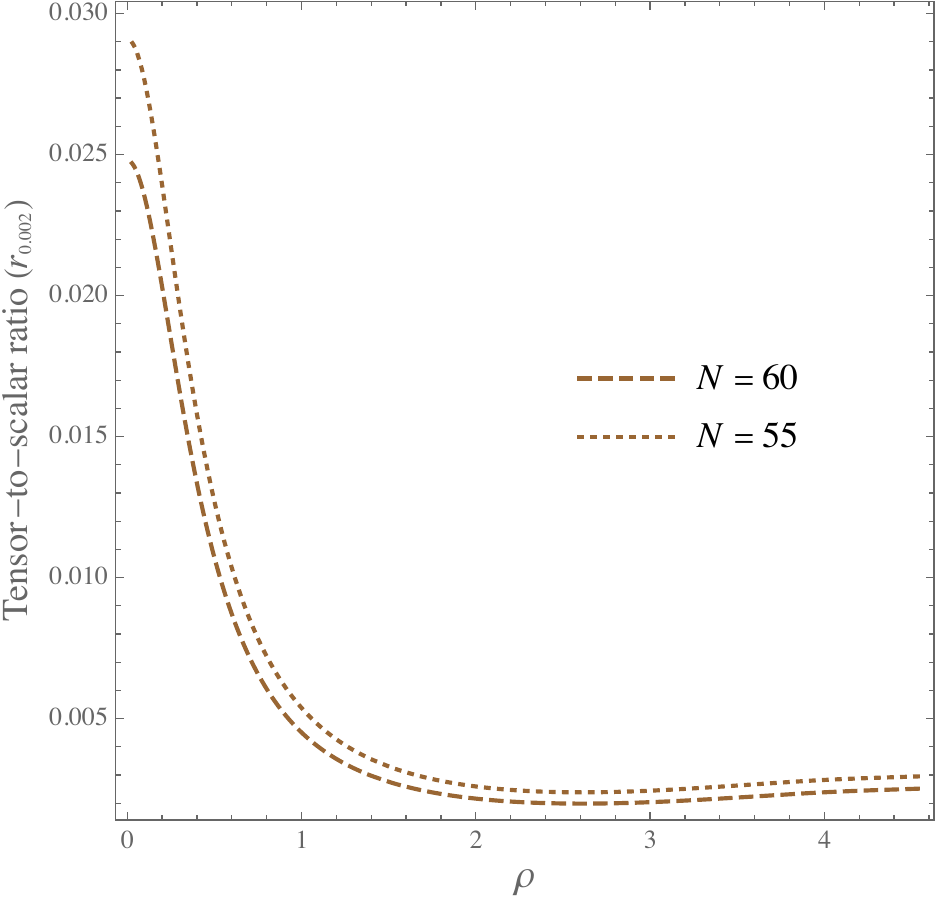} 
 \end{center}

   \caption{\em Analogous to Fig.~\ref{nsr}, but in the presence of a non-minimal coupling (Eq.~(\ref{FtN})).}
\label{nsr2}
\end{figure}

In Fig.~\ref{nsr2} we show the analogous plots but in the presence of a non-minimal coupling, which was chosen to be the one in Eq.~(\ref{FtN}). For the value of $\alpha$ considered in this figure ($\alpha=0.6>1/2$) there are two minima of the Einstein-frame potential in the band $\phi\in[\pi f, 2\pi f]$, as discussed in Sec.~\ref{Stationary points}. The inflatons always roll towards the global minimum for the values of the parameters and initial conditions considered in  Fig.~\ref{nsr2}. The largest value of $\beta_{\rm iso}$ found is again of order $10^{-4}$  and  obtained for the largest value of $\rho$. In this figure one can observe a qualitatively different dependence of $n_s$ and $r$ on $\rho$ (for large $\rho$) compared to  Fig.~\ref{nsr}. Moreover, we nicely recover the known result~\cite{Ferreira:2018nav,Reyimuaji:2020goi} that a non-minimal coupling alone can improve the agreement with cosmological data\footnote{Other known ways, which allow natural inflation to agree with cosmological data, are the inclusion of a Weyl-squared term~\cite{Salvio:2019wcp}, promoting the connection to an independent dynamical variable~\cite{Antoniadis:2018yfq} (Palatini formulation of gravity) and considering inflaton interactions with a gauge sector during inflation~\cite{Reyimuaji:2020bkm}.}.

\begin{figure}[t]
\begin{center}
 \includegraphics[scale=0.52]{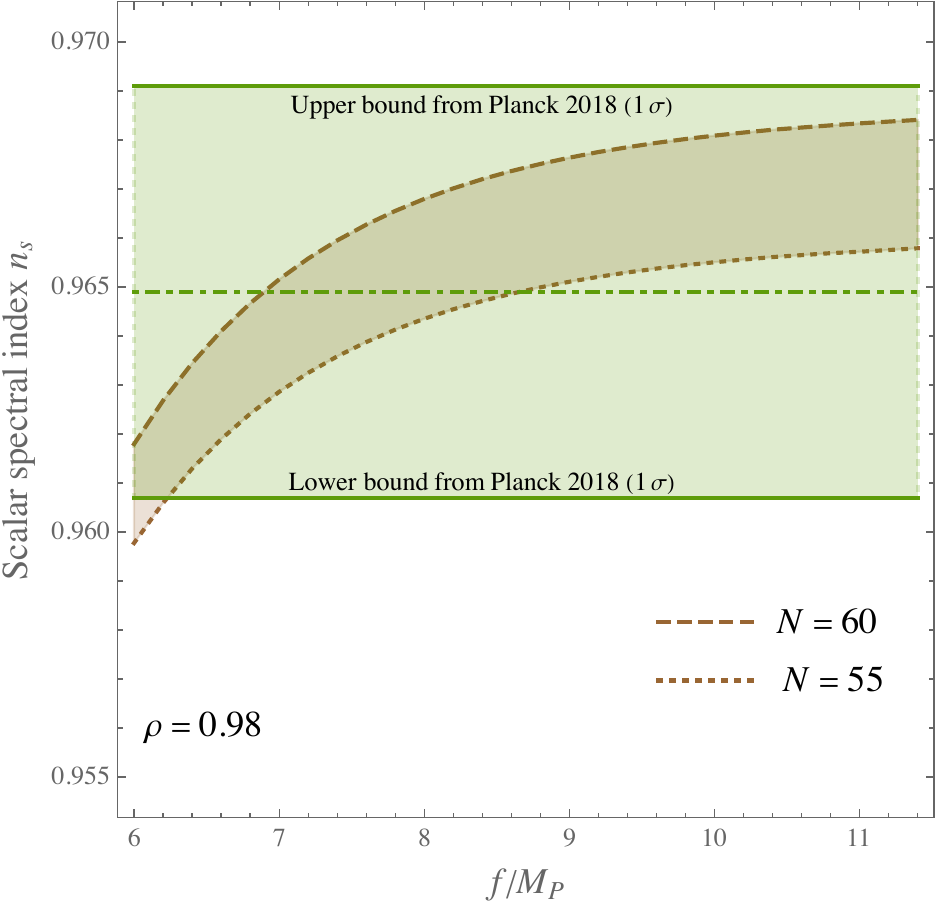}  
\hspace{0.6cm}  
\includegraphics[scale=0.52]{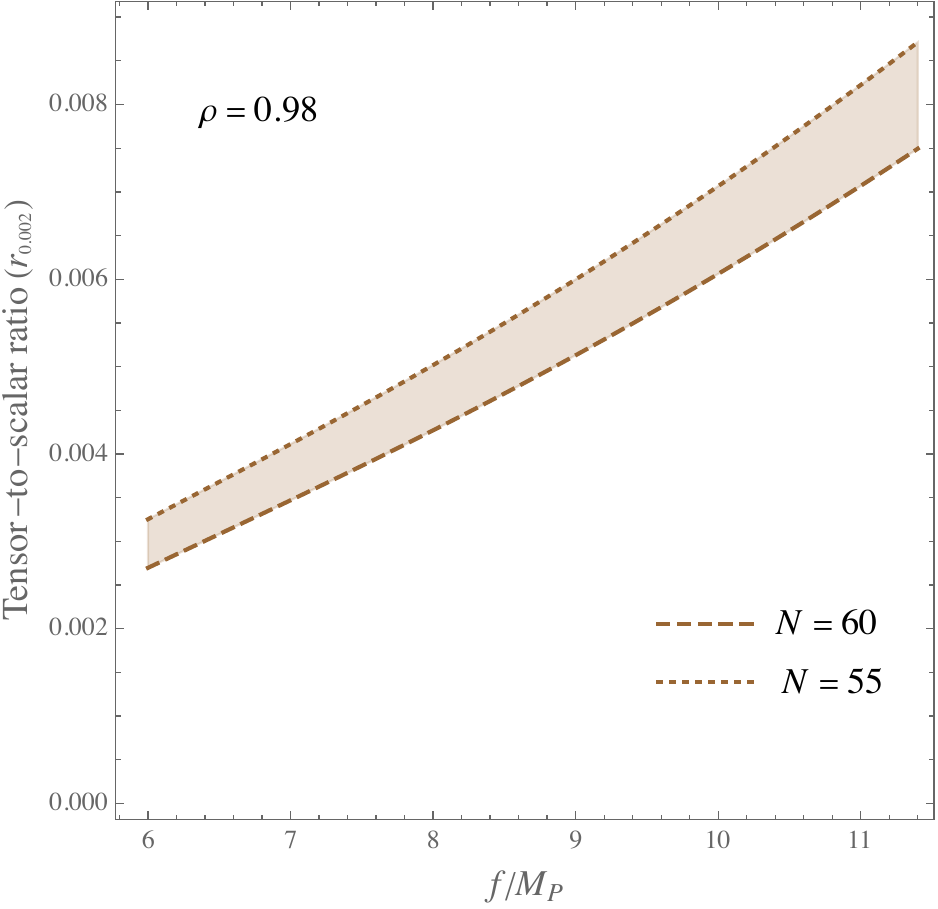} 
 \end{center}

   \caption{\em {\bf Left plot:} comparison between the theoretical predictions for $n_s$ as a function of  $f$ and the Planck observational bounds in~\cite{Ade:2015lrj}.  {\bf Right plot:} corresponding variation of $r_{0.002}$. }
   \vspace{0.1cm}
   {\em 
   In this figure a vanishing non-minimal coupling has been chosen, $F=\bp^2$.
   }
\label{fnsr}
\end{figure}

In Fig.~\ref{fnsr} the dependence of $n_s$ and $r$ on $f$ is shown for a fixed value of $\rho$ and in the absence of the non-minimal coupling. The $1\sigma$ bounds from Planck 2018~\cite{Ade:2015lrj} are explicitly shown in the left plot for $n_s$. Also here we find very small values of $\beta_{\rm iso}$: the largest value is of order $10^{-5}$ and is obtained for the smallest value of $f$ considered in  Fig.~\ref{fnsr}.

\begin{figure}[t]
\begin{center}
 \includegraphics[scale=0.52]{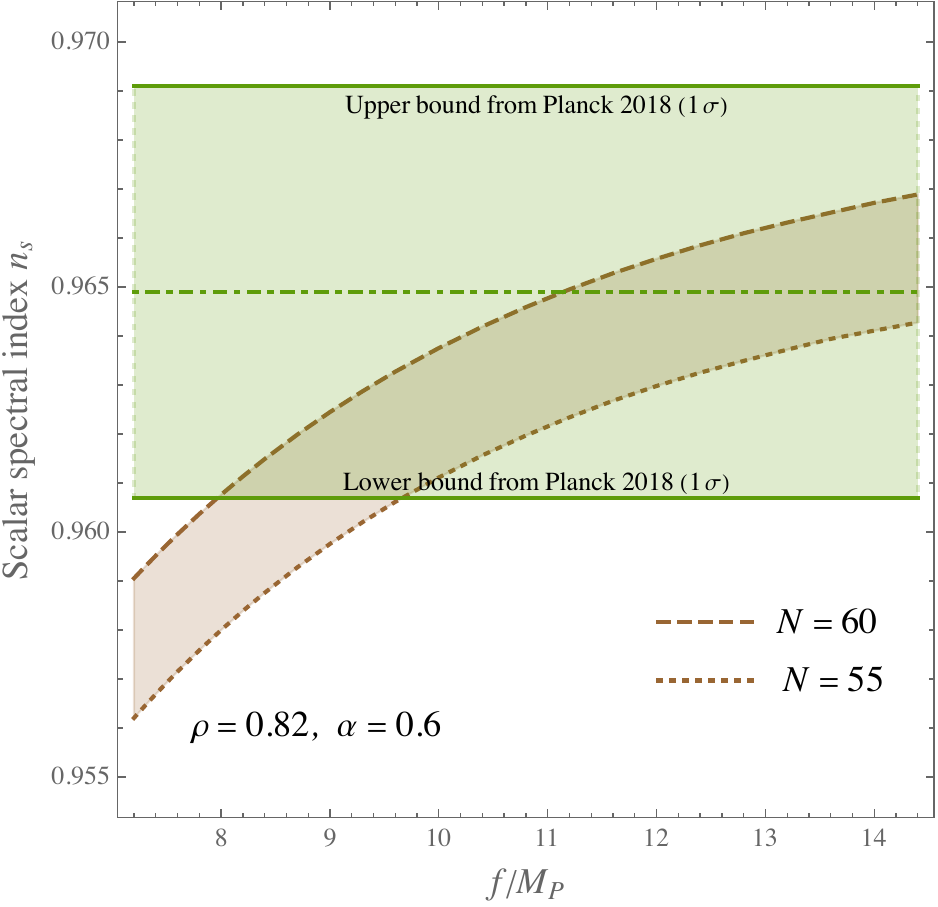}  
\hspace{0.6cm}  
\includegraphics[scale=0.52]{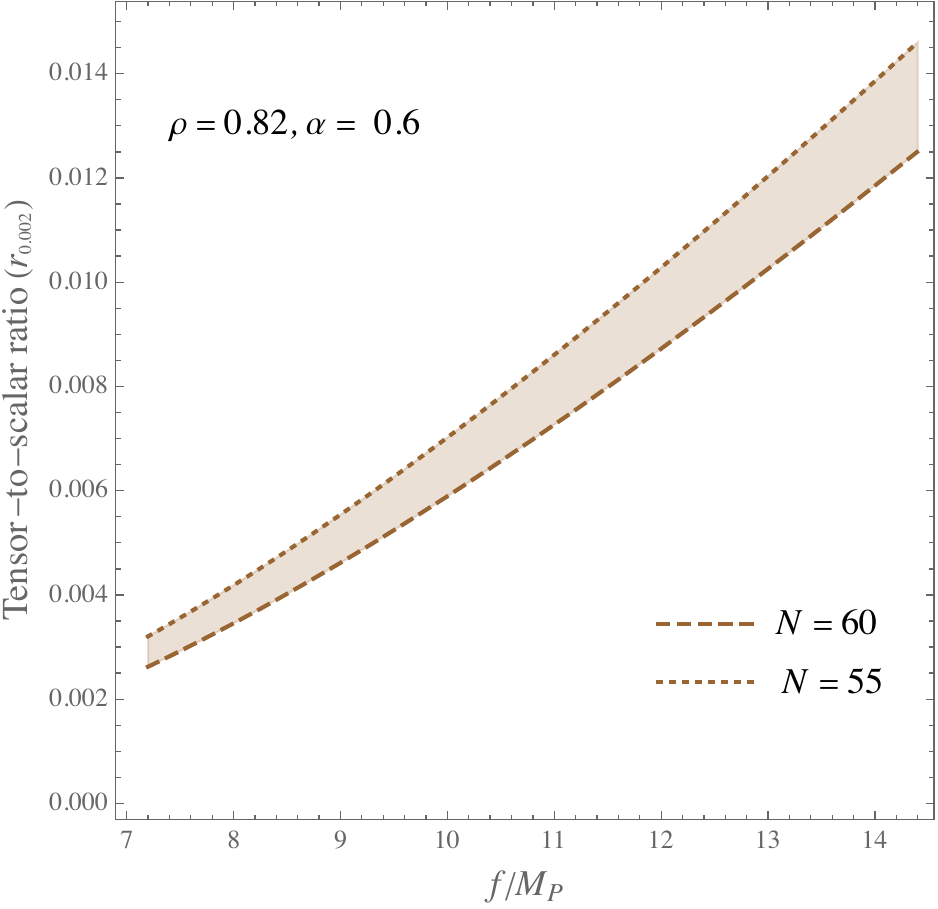} 
 \end{center}

   \caption{\em Analogous to Fig.~\ref{fnsr}, but in the presence of a non-minimal coupling (Eq.~(\ref{FtN})).}
\label{fnsr2}
\end{figure}

 In Fig.~\ref{fnsr2} we show the analogous plots but in the presence of a non-minimal coupling, which is chosen to be the one in Eq.~(\ref{FtN}). The same reference value $\alpha=0.6$ as in Fig.~\ref{nsr2} was chosen. The values of $\beta_{\rm iso}$ are even much smaller than those found in Fig.~\ref{fnsr}.

 \section{Conclusions}\label{Conclusions}

Let us provide a detailed summary of the paper.

\begin{itemize} 
\item In this work a new multifield inflationary scenario has been presented, where a PNGB $\phi$ (with a naturally flat potential) and the effective Starobinsky scalar $z$, the scalaron, are both active during inflation. The potential of $\phi$ is protected from quantum corrections, which, however, generically generate an $R^2$ term that is equivalent to the scalaron. Taking into account  the inflationary dynamics  of both $\phi$ and $z$ is thus well motivated. 
\item For the sake of generality a non-minimal coupling $F(\phi)$ between the PNGB and the Ricci scalar has also been included. Indeed, as shown explicitly in the appendix, both $V(\phi)$ and $F(\phi)$ generically emerge from quantum gravity effects (although $F$ can be close to the minimal value $\bp^2$ in some specific theories).
\item We have found that the Einstein-frame potential $U(\phi,z)$ can have (besides a global minimum) other stationary points and even other minima depending on the values of the parameters.
\item In any case a robust inflationary attractor, which effectively reduces the system to a quasi-single-field inflation, is present even for order one values of $\rho$ (that corresponds to comparable values of $m_\phi$ and $m_z$), as shown in Fig.~\ref{attractor}. This ensures that the most recent bounds on isocurvature modes presented by the Planck collaboration are satisfied. We have also numerically checked that one recovers natural (scalaron) inflation for small (large) values of $\rho$, namely for $m_\phi\ll m_z$ ($m_\phi\gg m_z$). Thus, this system  interpolates between two very well motivated inflationary models.
\item This natural-scalaron inflation is in excellent agreement with Planck constraints on $n_s$ and $r$ in a large region of the parameter space and for a large set of initial conditions, as shown in Figs.~\ref{attractor},~\ref{nsr},~\ref{nsr2},~\ref{fnsr} and~\ref{fnsr2}. This is the case even in the absence of non-minimal couplings, that is $F=\bp^2$ (see Figs.~\ref{attractor},~\ref{nsr} and~\ref{fnsr}). Therefore, natural-scalaron inflation is a well-motivated way of bringing natural inflation into perfect agreement with the observational bounds.
\end{itemize}

As a final remark, let us note that an interesting possible outlook would be to investigate whether the model presented and studied here can be tested with gravitational wave detectors. Some future gravitational wave space-borne interferometers  will be maybe able to provide extra tests in addition to those given by CMB observations, like in some versions of Higgs inflation~\cite{Salvio:2021kya}.

 \subsection*{Acknowledgments}
 I thank Marina Migliaccio for useful communications.

  \appendix
  \section{A possible microscopic origin of $V(\phi)$ and $F(\phi)$}\label{microrigin}

  In this appendix we illustrate a possible microscopic origin not only of the natural potential $V$, which has been briefly discussed in a number of articles (see e.g.~\cite{Freese:1990rb}), but also of the natural non-minimal coupling function $F$. To the best of our knowledge the microscopic origin of $F$ has not been discussed before\footnote{However, microscopic origins of non-minimal couplings for other types of scalars have already been discussed before, see e.g.~\cite{Barbon:2015fla} for a discussion on how a  Standard Model Higgs non-minimal coupling could emerge integrating out an additional scalar. }.
  
  The simplest possibility, which will be treated here, is to consider a version of QCD with a confinement scale $f$ around the Planck scale and with three flavors of tilde-quarks: $\tilde q=\{\tilde u, \tilde d, \tilde s\}\equiv \{\tilde q_1, \tilde q_2, \tilde q_3\}$, where the tilde distinguishes from the analogous QCD quantities.
 Like in ordinary QCD the strong dynamics forms condensates with a typical scale  $\tau$ such that~\cite{Weinberg2} 
\be \langle  \bar{\tilde q}'_i \tilde q'_j \rangle = -\tau \delta_{ij}, \qquad \langle\bar{\tilde q}'_i \gamma_5\tilde q'_j \rangle =0, \label{qVEV} \ee  
where $\langle\cdot\rangle$ represents the vacuum expectation value and $\tilde q'_i$  are the Goldstone-free quark fields:
\be \tilde q' = \exp(i \gamma_5 B/(\sqrt{2}f)) \tilde q. \label{qp}\ee 
 Also, $f$ is analogous to the pion decay constant and  $B$ is the Hermitian matrix containing the tilde-mesons (which have canonically normalized kinetic terms)
\be \label{Bf}   B\equiv \left(\baccc \frac{\tilde\pi^{0}}{\sqrt{2}} +\frac{\tilde\eta^0}{\sqrt{6}}& \tilde\pi^+ & \tilde K^+ \\
(\tilde\pi^+)^\dagger & -\frac{\tilde\pi^{0}}{\sqrt{2}} +\frac{\tilde\eta^0}{\sqrt{6}} & \tilde K^0 \\
(\tilde K^+)^\dagger & (\tilde K^0)^\dagger & -\sqrt{\frac{2}{3}}\tilde\eta^0 \ea \right). 
\ee
These scalars are the Goldstone bosons associated with the breaking of the axial part of the global ${\rm SU(3)}_{\rm f}$  flavor group (which rotates $\{\tilde u, \tilde d, \tilde s\}$). Just like in ordinary QCD, one can add quark mass terms such that the axial part of ${\rm SU(3)}_{\rm f}$  flavor group  is explicitly broken:
\be \mathscr{L}_{\rm mass} = -\bar{\tilde q} M_q\tilde q = -\bar{\tilde q}'  \exp(-i \gamma_5 B/(\sqrt{2}f)) M_q \exp(-i \gamma_5 B/(\sqrt{2}f)) \tilde q', \ee
where $M_q$ is the tilde-quark mass matrix. The group ${\rm SU(3)}_{\rm f}$ is an approximate symmetry as long as the elements of $M_q$ are small compared to $f$.
The tilde-meson potential can be computed  from these mass terms using~(\ref{qVEV}). 
In general this potential turns out to be equal to \be V = \tau\,  {\rm Tr}\left[ \cos\left(\sqrt{2}B/f\right) M_q\right] +\Lambda_0,\label{VNq}\ee
where $\Lambda_0$ is a ``bare"  cosmological constant term.

Using standard effective field theory methods~\cite{Weinberg2} and taking $M_q$ diagonal for simplicity,
\be M_q = {\rm diag}(m_{\tilde u},m_{\tilde d},m_{\tilde s}), \ee
one finds the following spectrum of the eight real PNGBs
\bea  m^2_{\tilde K^0} &=& \frac{\tau}{f^2}(m_{\tilde d} + m_{\tilde s}), \label{K0} \\ 
m^2_{\tilde K^+} &=& \frac{\tau}{f^2}(m_{\tilde u} + m_{\tilde s}), \label{Kp} \\ 
m^2_{\tilde \pi^0} &=& m^2_{\tilde \pi^+}  = \frac{\tau}{f^2}(m_{\tilde u} + m_{\tilde d}),  \label{pi0}\\
m^2_{\tilde \eta^0} &=& \frac{\tau}{f^2}\left(\frac{m_{\tilde u}+m_{\tilde d} + 4 m_{\tilde s}}{3}\right), \label{eta0}
\eea
%(See Sec. 19.7 of~\cite{Weinberg2} )
By using the known values of the meson masses, the up and down quark masses and the pion decay constant one obtains
% \xxx{I used that $\tau/f^3$ is not changed much by non-vanishing quark masses. This should be correct because $F_\pi \leftrightarrow f$ and $v\leftrightarrow \tau}$ (see  Sec. 19.7 of~\cite{Weinberg2}) are much bigger than the the up and down quark masses (which generate the pion mass) and the relation between $m^2_{\pi^0}$ and $m_u$ and $m_d$ is the same in the SU(2)$\times$SU(2) case and in the SU(3)$\times$ SU(3) case (see Sec.~19.7 of ~\cite{Weinberg2})}
\be \tau \sim 30 f^3 \label{kappaf}. \ee
This relation should also be approximately true in this version of QCD with $f$ taken at the Planck scale as long as the elements of $M_q$ are  much smaller than $f$.

 Now, by choosing an inverted hierarchy $m_{\tilde  u} \gg m_{\tilde d}, m_{\tilde  s}$ we obtain that the lightest pseudo-Goldstone boson is the complex scalar $\tilde K^0$.
During inflation  we can parameterize it as 
\be\tilde K^0 = \frac{\phi}{\sqrt{2}} \exp(i\alpha/f),\ee where $\phi$ is the real inflaton field we have introduced before and $\alpha$ is some angular real field. To compute the low energy potential for $\tilde K^0$ we can  set all other (very heavy) tilde-meson fields to zero in~(\ref{Bf}):
 \be \label{Bf2}   B= \left(\baccc 0& 0 &0 \\
0 & 0 &\frac{\phi}{\sqrt{2}}  \,  e^{i\alpha/f} \\
0 &\frac{\phi}{\sqrt{2}}  \, e^{-i\alpha/f} & 0 \ea \right). 
\ee
In this case the eigenvalues of $B$ are $0$,  $\phi/\sqrt{2}$ and $-\phi/\sqrt{2}$ so
 \be \label{cBf}  \cos\left(\frac{\sqrt{2}B}{f}\right)= P +\cos\left(\frac{\phi}{f}\right) (1-P)
\ee
where $P$ is the projector on the first (vanishing) eigenvalue of $B$, namely $P$ = diag$(1,0,0)$.
From~(\ref{VNq}) the potential is  given by
 \be V(\phi) = \tau (m_{\tilde d}+m_{\tilde s}) \cos\left(\frac{\phi}{f}\right) +\Lambda_0+\tau m_{\tilde u}.\ee
 Comparing this expression with~(\ref{VInf}) we obtain  $\Lambda = [\tau (m_{\tilde d}+m_{\tilde s})]^{1/4}$  and  $\Lambda_0 = \Lambda^4+\Lambda_{\rm cc}-\tau m_{\tilde u}$. Here we see how $\Lambda$ corresponds to the symmetry breaking parameters in the fundamental action (in this case the quark masses). Since $V$ is independent of $\alpha$,  slow-roll inflation occurs along trajectories of constant $\alpha$.
 
 The equations derived so far allow us to estimate the masses  of the lightest tilde-quarks and of the lightest tilde-meson, $m_\phi$.  When $\phi$ gives a dominant contribution to inflation the parameter $\Lambda$ is  below the Planck scale, around $10^{-2} \bp$ or $10^{-3} \bp$,
 %I used 6 \times 10^{-3}
  as a consequence of the observational constraint in~(\ref{PRobserved}). Using the expression of $m_\phi$ in~(\ref{mpmz}) and typical values of $f$ (see Fig~\ref{fnsr}) 
 %I used f=6.6 \bp$
  one obtains that $m_\phi$ is around the scale $10^{-5} \bp$. The relation in~(\ref{K0}) and~(\ref{kappaf}) then tell us (modulo large hierarchies between $m_{\tilde d}$ and $m_{\tilde s}$) that the masses of the lightest tilde-quarks are around $10^{-13} \bp\sim 10^5~$GeV. Note in particular that these mass scales are large enough to leave the standard late cosmology unaffected.
 
 Let us see now if  $\alpha$ satisfies the observational isocurvature bounds. The $\tilde K^0$ kinetic term in the effective Lagrangian  reads
 \be \frac{1}{2}\partial_\mu\phi\partial^\mu\phi+
 \frac{\phi^2}{2f^2}\partial_\mu\alpha\partial^\mu\alpha.\label{kinphia}\ee
 This kinetic term features a (flat) field metric written in a field coordinate system such that the field Christoffel symbols are not all zero: their non-zero components are 
 \be \Gamma^\phi_{\alpha\alpha} = -\frac{\phi}{f^2}, \quad \Gamma^\alpha_{\phi\alpha}=\Gamma^\alpha_{\alpha\phi} =\frac1{\phi}.\ee
 Therefore, the effective field-dependent scalar squared-mass matrix, whose elements are defined by $m^2_{ij} \equiv \nabla_i\nabla_j V$, where  $\nabla_i$ are the covariant derivatives on the field space computed with the metric in~(\ref{kinphia}), are 
 \be m_{\phi\phi}^2 = \partial_\phi^2 V, \quad m_{\alpha\alpha}^2 = -\Gamma^\phi_{\alpha\alpha} \partial_\phi V = \frac{\phi}{f^2} \partial_\phi V \ee
 and no mixing, i.e. $m_{\phi\alpha}=0$. Taking the potential  in~(\ref{VInf}) the explicit expressions are 
 \be m_{\phi\phi}^2 = -\frac{\Lambda ^4 \cos \left(\frac{\phi }{f}\right)}{f^2}, \qquad m_{\alpha\alpha}^2 =-\frac{\Lambda ^4 \phi  \sin \left(\frac{\phi }{f}\right)}{f^3}.\ee
 Since we focus on the interval $\phi\in [\pi f,2\pi f]$ we see that during the whole duration of inflation $m_{\alpha\alpha}^2 >0$. With this choice  one finds that for field values corresponding to about 60 e-folds before the end of inflation (and for $f\sim 10 \bp$)  the positive value of $m_{\alpha\alpha}^2$ is of order $H$. Therefore, following the formalism of~\cite{Kaiser:2012ak},   the isocurvature perturbation associated with $\alpha$ turns out to be highly suppressed in the superhorizon limit and the observational bounds are satisfied.

A sizable non-minimal coupling could appear due to non-renormalizable interactions between $\tilde q$ and gravity predicted at low energies by some theories of quantum gravity. It is interesting to illustrate how this can happen. Consider a quantum gravity scenario that leads to the effective low energy couplings 
\be \mathscr{L}_{qR}  = \frac{m_P^2}{2} R -\frac{1}{2\bp}\bar{\tilde q} J \tilde q R = \frac{m_P^2}{2} R-\frac{J_{ij}}{2\bp}\bar{\tilde q}_i \tilde q_j R, \label{LqRs}\ee
where $m_P^2 R/2$ is a ``bare" Einstein-Hilbert term and $J$ is a $3\times 3$ matrix of constant real coefficients. By using~(\ref{qVEV}) and~(\ref{qp}) one finds the following effective term proportional to $R$
\be \frac{m_P^2}{2} R+ \frac{\tau}{2\bp} {\rm Tr}\left[\cos\left(\sqrt{2}B/f\right) J\right] R,\ee
which, using~(\ref{cBf}), has the form $F(\phi)R/2$ with $F$ given in~(\ref{FtN}) and the identifications
\be   \alpha   = \frac{\tau}{M_P^3}\Tr(J-PJ), \qquad m_P^2 =   (1+\alpha)\bp^2-\frac{\tau}{M_P}\Tr(PJ). \ee
%\oo{
%A more general non-minimal coupling between $\phi$ and $R$ can emerge from a term of the form  
%\be \mathscr{L}_{qR}  =\mathscr{W}(\bar{\tilde q} J \tilde q) R, \ee
%where $\mathscr{W}$ is a generic function (in Eq.~(\ref{LqRs}) the special case $\mathscr{W}(x)=-x/(2\bp)$ has been considered). By using again~(\ref{qVEV}) and~(\ref{qp}) one finds the general effective non-minimal coupling
%\be  \mathscr{W}(-\tau{\rm Tr}\left[\cos\left(B/f\right) J\right])\, R.\ee
%From~(\ref{cBf}) it is clear that this term depends on $\phi$ only through the combination $\cos(\phi/f)$.
%}
%
%  
%  \section{Analytic expressions for the relevant inflationary quantities}
%  In the natural-scalaron inflationary scenario the slow-roll parameters are
%  \bea \epsilon &=&\nonumber \\  
%  \eta^{1}_{\,\,\, 1} &=&  \nonumber \\  
%  \eta^{2}_{\,\,\, 2} &=& \eea
%while the quantities $P_{\mathcal R}$ and $n_s$ and $r$ in~(\ref{power-spectrum}),~(\ref{nsFormula}) and~(\ref{rW}) read
%\bea P_{\mathcal R} &=& \frac{\left[(\partial_\phi N)^2+(\partial_zN)^2\right] \left[576 \beta V(\phi)
%+\left(z^2-6 F(\phi )\right)^2\right]}{1152 \pi ^2 \beta  z^2} \\ 
%n_s &=&, \\
%r&=& \frac{48}{z^2 \left[(\partial_\phi N)^2+(\partial_zN)^2\right]}\eea
%  
%  
 
 \section{Field-dependent covariant masses}\label{Field dependent covariant masses}
 
 Following the formalism of~\cite{Kaiser:2012ak} (see also~\cite{Gordon:2000hv} for a previous work with a flat field metric), the most important quantities to estimate the size of the isocurvature perturbations in the slow-roll approximation are the elements of the field-dependent covariant squared-mass matrix, $m_{ij}^2 \equiv \nabla_i\nabla_j U$, where  $\nabla_i$ are the covariant derivatives on the field space computed with the field metric $K_{ij}$ in~(\ref{action}). Explicitly, 
\be  m_{ij}^2 = (\partial_i \delta^k_j- \gamma^k_{ij})\partial_kU.\ee

For actions of the form~(\ref{Gammast}), with $i=1,2$ and $\phi^1 = \phi$ and $\phi^2 =z$,  the $m^2_{ij}$ are given by 
\bea m^2_{11}&=& \frac{3 \bp^4 \left[z^2 \left(6 F'^2+48 \beta  V''-z^2  F''\right)+2 z^2 F \left(3 F''-1\right)+12 F^2+192 \beta  V\right]}{4 \beta  z^6},\nonumber\\
m^2_{22} &=& \frac{3 \bp^4 \left(12 F^2+192 \beta  V-z^2 F\right)}{\beta  z^6}, \nonumber \\
m^2_{12} &=& m^2_{21}=\frac{3 \bp^4 \left[\left(z^2-18 F\right) F'-144 \beta  V'\right]}{4 \beta  z^5}. \nonumber \eea
These expressions hold for generic (and not necessarily periodic) $V$ and $F$.

At this point, it is convenient to introduce the unit vector $\hat\sigma^i$ tangent to the inflationary path $\phi^i=\phi_0^i$,
\be \hat\sigma^i\equiv \frac{\dot\phi_0^i}{\sqrt{K_{ij}\dot\phi_0^i\dot\phi_0^j}}, \ee
 and the set of unit vectors orthogonal to the inflationary path. In the presence of two inflatons
  we have only one of such orthogonal unit vectors,  $\hat s^i$ (see e.g.~\cite{Gundhi:2018wyz}) .  For actions of the form~(\ref{Gammast}), the explicit expression of $\hat s^i$ is
 \be  \hat s^1 \equiv  \frac{\dot z_0}{\sqrt{K_{ij}\dot\phi_0^i\dot\phi_0^j}}, \qquad \hat s^2 \equiv  \frac{-\dot \phi_0}{\sqrt{K_{ij}\dot\phi_0^i\dot\phi_0^j}}.\ee
The key quantities are in particular the projections of $m_{ij}^2$ on $\hat\sigma^i$ and $\hat s^i$:
\be m^2_{\sigma\sigma} \equiv \hat\sigma^i\hat\sigma^jm_{ij}^2,\qquad m^2_{ss} \equiv \hat s^i\hat s^jm_{ij}^2.  \ee
The effective  mass $m_{\sigma\sigma}$ corresponds to the usual curvature perturbations, while
   $m_{ss}$ corresponds to the isocurvature perturbations.

%\section{Conclusions}\label{Conclusions}

\vspace{0.7cm}

 \footnotesize
\begin{multicols}{2}

\end{multicols}


\begin{thebibliography}{}
\small{

%\cite{Freese:1990rb}
\bibitem{Freese:1990rb}
  K.~Freese, J.~A.~Frieman and A.~V.~Olinto,
  ``Natural inflation with pseudo - Nambu-Goldstone bosons,''
  Phys.\ Rev.\ Lett.\  {\bf 65} (1990) 3233.
F.~C.~Adams, J.~R.~Bond, K.~Freese, J.~A.~Frieman and A.~V.~Olinto,
  ``Natural inflation: Particle physics models, power law spectra for large scale structure, and constraints from COBE,''
  Phys.\ Rev.\ D {\bf 47} (1993) 426
  [\hhref{hep-ph/9207245}].

%\cite{Pajer:2013fsa}
\bibitem{Pajer:2013fsa}
E.~Pajer and M.~Peloso,
``A review of Axion Inflation in the era of Planck,''
Class. Quant. Grav. \textbf{30}, 214002 (2013)
[\hhref{1305.3557}].

%\cite{Utiyama:1962sn}
\bibitem{Utiyama:1962sn}
  R.~Utiyama and B.~S.~DeWitt,
  ``Renormalization of a classical gravitational field interacting with quantized matter fields,''
  J.\ Math.\ Phys.\  {\bf 3} (1962) 608.


  \bibitem{Starobinsky:1980te} 
  A.~A.~Starobinsky,
  ``A New Type of Isotropic Cosmological Models Without Singularity,''
  Phys.\ Lett.\ B {\bf 91}, 99 (1980).


 %\cite{Ade:2015lrj}
\bibitem{Ade:2015lrj}
  P.~A.~R.~Ade {\it et al.} [Planck Collaboration],
  ``Planck 2015 results. XX. Constraints on inflation,''
  Astron.\ Astrophys.\  {\bf 594} (2016) A20 
  [\hhref{1502.02114}].
  Y.~Akrami {\it et al.} [Planck Collaboration],
  ``Planck 2018 results. X. Constraints on inflation,''
  Astron. Astrophys. \textbf{641} (2020), A10
  [\hhref{1807.06211}].

%\cite{Kaneda:2015jma}
\bibitem{Kaneda:2015jma}
S.~Kaneda and S.~V.~Ketov,
``Starobinsky-like two-field inflation,''
Eur. Phys. J. C \textbf{76}, no.1, 26 (2016)
[\hhref{1510.03524}].
T.~Mori, K.~Kohri and J.~White,
``Multi-field effects in a simple extension of $R^2$ inflation,''
JCAP \textbf{10}, 044 (2017)
[\hhref{1705.05638}].
S.~Pi, Y.~l.~Zhang, Q.~G.~Huang and M.~Sasaki,
``Scalaron from $R^2$-gravity as a heavy field,''
JCAP \textbf{05}, 042 (2018)
[\hhref{1712.09896}].
D.~D.~Canko, I.~D.~Gialamas and G.~P.~Kodaxis,
``A simple $F(\mathcal{R},\phi )$ deformation of Starobinsky inflationary model,''
Eur. Phys. J. C \textbf{80}, no.5, 458 (2020)
[\hhref{1901.06296}].
A.~Gundhi, S.~V.~Ketov and C.~F.~Steinwachs,
``Primordial black hole dark matter in dilaton-extended two-field Starobinsky inflation,''
Phys. Rev. D \textbf{103}, no.8, 083518 (2021)
[\hhref{2011.05999}].

 \bibitem{Bezrukov:2007ep}
  F.~L.~Bezrukov and M.~Shaposhnikov,
  ``The Standard Model Higgs boson as the inflaton,''
  Phys.\ Lett.\ B {\bf 659} (2008) 703
  [\hhref{0710.3755}].
  
%\cite{Salvio:2015kka}
\bibitem{Salvio:2015kka}
  A.~Salvio and A.~Mazumdar,
  ``Classical and Quantum Initial Conditions for Higgs Inflation,''
  Phys.\ Lett.\ B {\bf 750} (2015) 194
  [\hhref{1506.07520}].
  X.~Calmet and I.~Kuntz,
``Higgs Starobinsky Inflation,''
Eur. Phys. J. C \textbf{76}, no.5, 289 (2016)
[\hhref{1605.02236}]. 


%\cite{Salvio:2016vxi}
\bibitem{Salvio:2016vxi}
  A.~Salvio,
  ``Solving the Standard Model Problems in Softened Gravity,''
  Phys.\ Rev.\ D {\bf 94} (2016) no.9,  096007
  [\hhref{1608.01194}].
  %%CITATION = doi:10.1103/PhysRevD.94.096007;%%
  
\bibitem{Ema}
Y.~C.~Wang and T.~Wang,
``Primordial perturbations generated by Higgs field and $R^2$ operator,''
Phys. Rev. D \textbf{96}, no.12, 123506 (2017)
[\hhref{1701.06636}].
Y.~Ema,
``Higgs Scalaron Mixed Inflation,''
Phys. Lett. B \textbf{770}, 403-411 (2017)
[\hhref{1701.07665}].
M.~He, A.~A.~Starobinsky and J.~Yokoyama,
``Inflation in the mixed Higgs-$R^2$ model,''
JCAP \textbf{05}, 064 (2018)
[\hhref{1804.00409}].



     %\cite{Gundhi:2018wyz}
\bibitem{Gundhi:2018wyz}
A.~Gundhi and C.~F.~Steinwachs,
``Scalaron-Higgs inflation,''
Nucl. Phys. B \textbf{954}, 114989 (2020)
[\hhref{1810.10546}].


\bibitem{Enckell}
V.~M.~Enckell, K.~Enqvist, S.~Rasanen and L.~P.~Wahlman,
``Higgs-$R^2$ inflation - full slow-roll study at tree-level,''
JCAP \textbf{01}, 041 (2020)
[\hhref{1812.08754}].
A.~Gundhi and C.~F.~Steinwachs,
``Scalaron-Higgs inflation reloaded: Higgs-dependent scalaron mass and primordial black hole dark matter,''
Eur. Phys. J. C \textbf{81}, no.5, 460 (2021)
[\hhref{2011.09485}].

       %\cite{Kannike:2015apa}
\bibitem{Kannike:2015apa}
A.~Salvio and A.~Strumia,
``Agravity,''
JHEP \textbf{06}, 080 (2014)
[\hhref{1403.4226}].
  K.~Kannike, G.~Hutsi, L.~Pizza, A.~Racioppi, M.~Raidal, A.~Salvio and A.~Strumia,
  ``Dynamically Induced Planck Scale and Inflation,''
  JHEP {\bf 1505} (2015) 065
   [\hhref{1502.01334}]. 
   
   \bibitem{JKubo}
J.~Kubo, M.~Lindner, K.~Schmitz and M.~Yamada,
``Planck mass and inflation as consequences of dynamically broken scale invariance,''
Phys. Rev. D \textbf{100}, no.1, 015037 (2019)
[\hhref{1811.05950}].


%\cite{Salvio:2020axm}
\bibitem{Salvio:2020axm}
A.~Salvio,
``Dimensional Transmutation in Gravity and Cosmology,'' 
Int. J. Mod. Phys. A \textbf{36} (2021) no.08n09, 2130006
[\hhref{2012.11608}].

%\cite{McDonough:2020gmn}
\bibitem{McDonough:2020gmn}
E.~McDonough, A.~H.~Guth and D.~I.~Kaiser, ``Nonminimal Couplings and the Forgotten Field of Axion Inflation,''
[\hhref{2010.04179}].

 %\cite{Stelle:1976gc}
\bibitem{Stelle:1976gc}
  K.~S.~Stelle,
  ``Renormalization of Higher Derivative Quantum Gravity,''
  Phys.\ Rev.\ D {\bf 16} (1977) 953.



  %\cite{Salvio:2018crh}
\bibitem{Salvio:2018crh}
A.~Salvio,
``Quadratic Gravity,''
Front. in Phys. \textbf{6} (2018), 77
[\hhref{1804.09944}].


%\cite{Salvio:2019wcp}
\bibitem{Salvio:2019wcp}
A.~Salvio,
``Quasi-Conformal Models and the Early Universe,''
Eur. Phys. J. C \textbf{79} (2019) no.9, 750
[\hhref{1907.00983}].



%\cite{Ferreira:2018nav}
\bibitem{Ferreira:2018nav}
  R.~Z.~Ferreira, A.~Notari and G.~Simeon,
  ``Natural Inflation with a periodic non-minimal coupling,''
  JCAP {\bf 1811} (2018) no.11,  021
  [\hhref{1806.05511}].
  G.~Simeon,
``Scalar-tensor extension of Natural Inflation,''
JCAP \textbf{07}, 028 (2020)
[\hhref{2002.07625}].
  

  
%\cite{Chiba:2008rp}
\bibitem{Chiba:2008rp}
  T.~Chiba and M.~Yamaguchi,
  ``Extended Slow-Roll Conditions and Primordial Fluctuations: Multiple Scalar Fields and Generalized Gravity,''
  JCAP {\bf 0901} (2009) 019
  [\hhref{0810.5387}].
  %%CITATION = ARXIV:0810.5387;%%
  
 %\cite{Sasaki:1995aw}
\bibitem{Sasaki:1995aw}
  M.~Sasaki and E.~D.~Stewart,
  ``A General analytic formula for the spectral index of the density perturbations produced during inflation,''
  Prog.\ Theor.\ Phys.\  {\bf 95} (1996) 71
  [\hhref{astro-ph/9507001}].
  
  

 %\cite{Kaiser:2012ak}
\bibitem{Kaiser:2012ak}
D.~I.~Kaiser, E.~A.~Mazenc and E.~I.~Sfakianakis,
``Primordial Bispectrum from Multifield Inflation with Nonminimal Couplings,''
Phys. Rev. D \textbf{87}, 064004 (2013)
[\hhref{1210.7487}].

%\cite{Stein:2021uge}
\bibitem{Stein:2021uge}
N.~K.~Stein and W.~H.~Kinney,
``Natural Inflation After Planck 2018,''
[\hhref{2106.02089}].

%\cite{Reyimuaji:2020goi}
\bibitem{Reyimuaji:2020goi}
Y.~Reyimuaji and X.~Zhang,
``Natural inflation with a nonminimal coupling to gravity,''
JCAP \textbf{03}, 059 (2021)
[\hhref{2012.14248}].

%\cite{Antoniadis:2018yfq}
\bibitem{Antoniadis:2018yfq}
I.~Antoniadis, A.~Karam, A.~Lykkas, T.~Pappas and K.~Tamvakis,
``Rescuing Quartic and Natural Inflation in the Palatini Formalism,''
JCAP \textbf{03}, 005 (2019)
[\hhref{1812.00847}].

%\cite{Reyimuaji:2020bkm}
\bibitem{Reyimuaji:2020bkm}
Y.~Reyimuaji and X.~Zhang,
``Warm-assisted natural inflation,''
JCAP \textbf{04}, 077 (2021)
[\hhref{2012.07329}].

%\cite{Barbon:2015fla}
\bibitem{Barbon:2015fla}
J.~L.~F.~Barbon, J.~A.~Casas, J.~Elias-Miro and J.~R.~Espinosa,
``Higgs Inflation as a Mirage,''
JHEP \textbf{09} (2015), 027
[\hhref{1501.02231}].

\bibitem{Weinberg2}
 S.~Weinberg ``The quantum theory of fields. Vol. 2: Modern applications". Cambridge University Press 1996.
 
 
%\cite{Gordon:2000hv}
\bibitem{Gordon:2000hv}
C.~Gordon, D.~Wands, B.~A.~Bassett and R.~Maartens,
``Adiabatic and entropy perturbations from inflation,''
Phys. Rev. D \textbf{63}, 023506 (2000)
[\hhref{astro-ph/0009131}].




  
  
%     
%  
%        %\cite{Bezrukov:2009db}
%\bibitem{Bezrukov:2009db}
%F.~L.~Bezrukov, A.~Magnin and M.~Shaposhnikov,
%  ``Standard Model Higgs boson mass from inflation,''
%  Phys.\ Lett.\ B {\bf 675} (2009) 88
%  [\hhref{0812.4950]}. 
%   %%CITATION = ARXIV:0812.4950;%%
%   
%   
%   \bibitem{Bezrukov:2009-2}
%  F.~Bezrukov and M.~Shaposhnikov,
%  ``Standard Model Higgs boson mass from inflation: Two loop analysis,''
%  JHEP {\bf 0907} (2009) 089
%  [\hhref{0904.1537}].
%  %%CITATION = ARXIV:0904.1537;%%
%  
%        \bibitem{Salvio-inf}
%    A.~Salvio,
%  ``Higgs Inflation at NNLO after the Boson Discovery,''
%  Phys.\ Lett.\ B {\bf 727} (2013) 234
%  [\hhref{1308.2244}].
%  
   

%     
%  %\cite{Ema:2019fdd}
%\bibitem{Ema:2019fdd}
%Y.~Ema,
%``Dynamical Emergence of Scalaron in Higgs Inflation,''
%JCAP \textbf{09} (2019), 027
%[\hhref{1907.00993}].
%
%%\cite{Calmet:2016fsr}
%\bibitem{Calmet:2016fsr}
%X.~Calmet and I.~Kuntz,
%``Higgs Starobinsky Inflation,''
%Eur. Phys. J. C \textbf{76} (2016) no.5, 289
%[\hhref{1605.02236}].
%
%
%   

%   \bibitem{Salvio:2017oyf}
%A.~Salvio,
%``Initial Conditions for Critical Higgs Inflation,''
%Phys. Lett. B \textbf{780} (2018), 111-117
%[\hhref{1712.04477}].
%
%\bibitem{Salvio:2018rv}
%A.~Salvio,
%``Critical Higgs inflation in a Viable Motivated Model,''
%Phys. Rev. D \textbf{99} (2019) no.1, 015037
%[\hhref{1810.00792}].

 
%  
%  %\cite{Bezrukov:2014bra}
%\bibitem{Bezrukov:2014bra}
%  F.~Bezrukov and M.~Shaposhnikov,
%  ``Higgs inflation at the critical point,''
%  Phys.\ Lett.\ B {\bf 734} (2014) 249
%  [\hhref{1403.6078}].
%  %%CITATION = ARXIV:1403.6078;%%
%
%  %\cite{Hamada:2014wna}
%\bibitem{Hamada:2014wna}
%  Y.~Hamada, H.~Kawai, K.~y.~Oda and S.~C.~Park,
%  ``Higgs inflation from Standard Model criticality,
%  Phys.\ Rev.\ D {\bf 91} (2015) 053008
%  [\hhref{1408.4864}].
%
%
%%\cite{Buttazzo:2013uya}
%\bibitem{Buttazzo:2013uya}
%D.~Buttazzo, G.~Degrassi, P.~P.~Giardino, G.~F.~Giudice, F.~Sala, A.~Salvio and A.~Strumia,
%``Investigating the near-criticality of the Higgs boson,''
%JHEP \textbf{12} (2013), 089
%[\hhref{1307.3536}].
%

 



 

 







%\cite{Aghanim:2018eyx}
\bibitem{Aghanim:2018eyx}
N.~Aghanim \textit{et al.} [Planck],
``Planck 2018 results. VI. Cosmological parameters,''
Astron. Astrophys. \textbf{641} (2020), A6
[\hhref{1807.06209}].

%\cite{Salvio:2021kya}
\bibitem{Salvio:2021kya}
A.~Salvio,
``Hearing Higgs with gravitational wave detectors,''
JCAP \textbf{06}, 040 (2021)
[\hhref{2104.12783}].


     }
  
  
 
  
  
 







\end{thebibliography}
\end{document}